\documentclass[preprint,11pt]{elsarticle}
\usepackage[utf8]{inputenc}
\usepackage{caption}
\usepackage{subcaption}
\usepackage{amssymb}
\usepackage[tbtags]{amsmath}
\usepackage{mathtools}
\usepackage{hyperref}
\usepackage{geometry}
\usepackage{booktabs}
\usepackage[frozencache,cachedir=.]{minted}
\usepackage{xcolor} 
\definecolor{LightGray}{gray}{0.9}

\hypersetup{colorlinks=true, linkcolor=blue}

\date{\today}

\begin{document}

\begin{frontmatter}

\title{Vacillating about media bias: changing one's mind intermittently within a network of political allies and opponents}


\journal{Physica A}

\author[add1]{Nicholas Low Kah Yean\corref{mycorrespondingauthor}}
\cortext[mycorrespondingauthor]{Corresponding author}
\ead{nlow1@student.unimelb.edu.au}

\author[add1,add2]{Andrew Melatos} 
\ead{amelatos@unimelb.edu.au}

\date{\today}
\address[add1]{School of Physics, University of Melbourne, Parkville, VIC 3010, Australia}
\address[add2]{Australian Research Council Centre of Excellence for Gravitational Wave Discovery (OzGrav), University of Melbourne, Parkville, VIC 3010, Australia}

\begin{abstract}
    One form of long-term behavior revealed by opinion dynamics simulations is intermittency, where an individual cycles between eras of stable, constant beliefs and turbulent, fluctuating beliefs, for example when inferring the political bias of a media organization. We explore this phenomenon by building an idealized network of Bayesian learners, who infer the bias of a coin from observations of coin tosses and peer pressure from political allies and opponents. Numerical simulations reveal that three types of network structure lead to three different types of intermittency, which are caused by agents ``locking out'' opponents from sure beliefs in specific ways. The probability density functions of the dwell times, over which the learners sustain stable or turbulent beliefs, differ across the three types of intermittency. Hence, one can observe the dwell times of a learner to infer the underlying network structure, at least in principle.
\end{abstract}


\end{frontmatter}

\newcommand{\mean}{\langle \theta \rangle}
\newcommand{\meanSub}[1]{\langle \theta \rangle_{#1}}
\newcommand{\stddev}{\sqrt{\langle ( \theta - \mean ) ^2 \rangle}}
\newcommand{\stddevSub}[1]{\sqrt{\langle ( \theta - \mean ) ^2 \rangle}_{#1}}
\newcommand{\asympTime}{t_{\rm a}}
\newcommand{\stableDwell}{t_{\rm s}}
\newcommand{\turbulentDwell}{t_{\rm t}}
\newcommand{\stablePDF}{p(\stableDwell)}
\newcommand{\turbulentPDF}{p(\turbulentDwell)}
\newcommand{\stableMean}{\langle \stableDwell \rangle}
\newcommand{\turbulentMean}{\langle \turbulentDwell \rangle}
\newcommand{\numInt}{n_{\rm int}}
\newcommand{\barabasi}{Barabási–Albert }
\newcommand{\erdos}{Erdős–Rényi }

\section{Introduction}

Media bias can alter behavioral patterns, such as voting in elections, by slanting the political opinions of media consumers \cite{druckman2005impact, eberl2017}. Granted that the opinions of media consumers can be skewed by media organizations, it is interesting to quantify how stable political opinions are. For example, what is the characteristic time-scale over which they remain unchanged? Furthermore, granted that opinions do change, it is interesting to identify the conditions under which this happens. For example, what role do network effects play, as opposed to the publication strategy of media organizations or the psychology of individual consumers? Substantial effort has been invested towards answering the question of whether the opinion held by an individual on a topic is prone to change (see Ref. \cite{opinionInstability} for an overview of opinion stability). Some studies find that political opinions display temporal fluctuations often \cite{converse1964, hill_2001}, whereas others find that political opinions are relatively stable over time \cite{ansolabehere_rodden_snyder_2008, druckman_fein_leeper_2012}. There are criticisms that measurements of opinion stability are inaccurate due to ambiguous survey questions or response categories that do not capture the respondents' true opinion \cite{ansolabehere_rodden_snyder_2008, achen_1975}. Some studies investigate the relationship between external messages (such as public debates \cite{hill_2001}, which showcase competing messages \cite{chong_druckman_2010}) and opinion change triggered by the psychology of individuals. However, few studies explicitly look at the influence of interactions with other people, where opinions change through peer pressure in a network, and the psychology of individuals is not the sole cause. One exception is Hoffman et al. \cite{hoffman2007}, who correlated the frequency that survey respondents discuss a political issue with others against the opinion stability of the respondents.

Complementing social science research, the field of opinion dynamics studies mathematically how human opinions diffuse and evolve in an idealized network of interacting agents \cite{xia2011multidiscipline}. Usually, models in opinion dynamics assume that the belief of an agent is sure; that is, it is described by a single-valued number\footnote{In the idealized example of a biased coin, which this paper investigates, an agent's belief is described as ``sure'', if the agent believes absolutely in one value of the coin's bias with 100\% conviction, or ``unsure", if the agent believes partially in several values of the coin's bias with less than 100\% conviction about each one. The adjectives ``sure'' and ``unsure'' are preferred instead of ``certain'' and ``uncertain'' to avoid the ambiguities caused by other meanings of the adjective ``certain'', e.g.\ particular.} \cite{noorazar2020recent}. A few models treat opinions as unsure; that is, they are described in a Bayesian framework by probability distributions, which may be multimodal for example \cite{jadbabaie2012non, fang2019social}. When there is antagonism between some agents, deterministic models with sure beliefs predict that the opinions of some agents are temporally unstable \cite{martins2010mass, shi2016evolution}, which may inhibit the agents from accurately inferring the bias of a media organization. In a previous paper \cite{nlow1}, we demonstrate that the same is true for probabilistic models with unsure beliefs: a network of Bayesian learners containing a mixture of political allies and opponents can display the counter-intuitive phenomenon of long-term intermittency, in which one or more learners cycle between eras of stability, where their beliefs are constant for many time steps, and eras of turbulence, where their beliefs fluctuate from one time step to the next. 

In this paper, we explore long-term intermittency in detail, as it is still relatively new in the literature. The goals are to (i) identify what conditions cause the phenomenon, and (ii) quantify its properties in preparation for future comparisons with real-world data. In Section \ref{sec:model}, we summarize the structure and update rule of the network of Bayesian learners, and describe the types of long-term behaviors of the networked agents, one of which is long-term intermittency. In Section \ref{sec:conditions}, we investigate what kind of network structure yields long-term intermittency and what proportion of agents experience intermittency. In Section \ref{sec:dwell_time_stats}, we compute the dwell time statistics of long-term intermittency, i.e.\ how many time steps an agent stays in the eras of stability or turbulence.

\section{Inferring media bias in a network}
\label{sec:model}

In this paper, as in previous work \cite{nlow1}, we map the difficult real-world problem of inferring media bias onto the idealized problem of inferring the bias of a coin. A network of $n$ agents strives to infer the hidden bias of a coin $0 \leq \theta_0 \leq 1$, defined as the probability that the coin lands heads after one toss. The beliefs of agent $i$ at time $t$ are unsure; they are described by a probability density function (PDF) $x_i(t,\theta)$, which can be multi-modal, and whose evolution therefore differs in important respects from sure beliefs defined by a single number \cite{degroot1974reaching, deffuant2000mixing, hegselmann2002opinion, martins2010mass, shi2016evolution}. At each discrete time step, the beliefs of the agents are shaped by two signals: (i) the external signal, which is a single coin toss, and (ii) the internal signals, which are the peer-pressure interactions of the agents with their political allies and opponents. In this section, we define a two-step Bayesian update rule, which encodes (i) and (ii), and summarize the three main types of long-term behavior it produces. Details of the model are expounded fully in Ref. \cite{nlow1}. We also study briefly the behavior of a one-step Bayesian update rule encoding (i) but not (ii) in \ref{sec:bayes_detailed}, in order to establish a baseline for network comparisons and develop intuition about how Bayes's Rule operates in the context of idealized opinion dynamics models of media bias.

\subsection{Idealized problem of inferring the bias of a coin: Bayesian update rule}
\label{sec:update_rules}

We assume that the outcome of the coin toss is an unfiltered public signal and is observed noiselessly by every agent in the network. At time $t$, this public signal $S(t) \in \{\text{heads}, \text{tails}\}$ updates the probabilistic beliefs of agent $i$ about the coin bias $0 \leq \theta \leq 1$ according to Bayes's rule:

\begin{equation}
\label{eq:bayes}
    x'_i(t+1/2,\theta) = \frac{P[S(t)|\theta]}{P[S(t)]}  x_i(t,\theta).
\end{equation}

\noindent In (\ref{eq:bayes}), $P[S(t)|\theta]$ is the likelihood, $x_i(t,\theta)$ is the prior probability distribution at time $t$, $x'_i(t+1/2,\theta)$ is the posterior probability distribution and $P[S(t)]=\sum_\theta P[S(t)|\theta] x_i(t,\theta)$ is the normalizing constant. Explicitly, the likelihood takes the following form:

\begin{equation}
\label{eq:actual_bayes}
    P[S(t)|\theta] =
    \begin{dcases}
        \theta,& \quad \text{if } S(t)  \text{ is heads} \\
        1-\theta,& \quad \text{if } S(t)  \text{ is tails.}
    \end{dcases}
\end{equation}

\noindent \ref{sec:bayes_detailed} explains how Bayes's rule relates to inferring the bias of a coin (and media bias more generally) by justifying the elements of Equations (1) and (2) and solving analytically the illustrative special case, where there is no peer pressure among  agents in the network.

Equation (\ref{eq:bayes}) is the first half of a two-step update rule, which is emphasized by writing the argument of $x'$ as $t+1/2$ instead of $t+1$. The second half involves network interactions, where the beliefs of the agents are further molded by peer pressure from their political allies and opponents. We encode the political relationship between agents $i$ and $j$ in the $n \times n$ adjacency matrix $A_{ij}$. For simplicity, we only consider three possible values of $A_{ij}$, $+1,-1,0$, which indicates that agents $i$ and $j$ are allies, opponents or strangers respectively. We also assume that $A_{ij}$ is a symmetric matrix; if agent $i$ perceives agent $j$ as an ally, then agent $j$ perceives agent $i$ as an ally. The posterior probability is updated according to the rule

\begin{equation}
\label{eq:update}
    x_i(t+1, \theta) \propto \max \left[0, \; x'_i(t + 1/2,\theta) + \mu \Delta x'_i(t + 1/2, \theta) \right].
\end{equation}

\noindent The proportionality constant in (\ref{eq:update}) is set via the condition $\sum_\theta x_i(t+1,\theta) = 1$. The learning rate $0.0<\mu\leq 0.5$ quantifies the susceptibility of an agent to the beliefs of its allies and opponents, and sets the time scale of the model. The bounds on $\mu$ are imposed to ensure that if agents 1 and 2 are allies and agent 1 is more confident in some $\theta_1$ compared to agent 2, then agent 1 remains more confident than agent 2 in $\theta_1$ after interacting \cite{nlow1}. The displacement

\begin{equation}
\label{eq:update_diff}
    \Delta x'_i(t+1/2, \theta) = 
        a_i^{-1} \sum_{j \neq i} A_{ij} \left[ x'_j(t+1/2,\theta) - x'_i(t+1/2,\theta) \right]
\end{equation}

\noindent with

\begin{equation}
\label{eq:a_i_def}
    a_i = \sum_{j \neq i} \left| A_{ij} \right|,
\end{equation}

\noindent quantifies the average difference in belief between agent $i$ and its allies and opponents. We use the signed rather than absolute difference in Equation (\ref{eq:update_diff}) to match the update rule in classic deterministic models, e.g. Deffuant-Weisbuch \cite{deffuant2000mixing}, and to avoid counterintuitive outcomes under certain circumstances.\footnote{By way of illustration, consider the following toy scenario that produces a counterintuitive outcome in a two-ally network. If one starts with $x_1(t=1/2,\theta=0.60) = 1.0$, $x_2(t=1/2,\theta=0.55) = x_2(t=1/2,\theta=0.60) = 0.5$, and $\mu=0.25$, then one obtains $x_2(t=1,\theta=0.55) = x_2(t=1,\theta=0.60) = 0.5$. That is, after agent 2 interacts with agent 1, agent 2's beliefs remain unchanged, whereas normal intuition would suggest that agent 2's confidence in $\theta=0.55$ and $\theta=0.60$ should decrease and increase respectively.} Iterating Equations (\ref{eq:bayes})--(\ref{eq:a_i_def}) yields one full time step from $t$ to $t+1$.

For simplicity, we make a few assumptions about the parameters of the model. Unless stated otherwise, we choose units that give $\mu=0.25$ and $\theta_0=0.6$ for ease of comparison with Ref. \cite{nlow1}. For computational efficiency, and because there is a limit to how fine-grained human opinions are in reality, we grid the continuous variable $\theta$ into 21 regularly-spaced values, i.e.\ $\theta = \{0.00, 0.05, ..., 1.00\}$. The priors, $x_i(t=0,\theta)$, are independently generated by sampling a Gaussian with mean and standard deviation in the ranges $[0.0, 1.0]$ and $[0.1, 0.4]$ respectively, discretized and truncated to the domain $0 \leq \theta \leq 1$. \ref{sec:discretization_visual} visualizes the discretization of $\theta$ and how the priors are generated.

\subsection{Relation to alternative update rules}
\label{sec:alternative_update_rules}

The update rule presented in this paper resembles the update rules in some deterministic models of opinion dynamics in the literature. Adjusting the opinion of an agent based on the difference in opinion between that agent and its allies and opponents through Equation (\ref{eq:update}) is similar to the Deffuant-Weisbuch model \cite{deffuant2000mixing} and extensions to it that incorporate antagonism \cite{jager2005uniformity, huet2010openness}, although the Deffuant-Weisbuch model is not probabilistic. The update rule of the deterministic Deffuant-Weisbuch model for two allied agents 1 and 2 has the following form:

\begin{align}
    x_1(t+1) &= x_1(t) + \mu \left[x_2(t) - x_1(t) \right] \label{eq:dw1} \\
    x_2(t+1) &= x_2(t) + \mu \left[x_1(t) - x_2(t) \right] \label{eq:dw2}.
\end{align}

\noindent Equations (\ref{eq:update})--(\ref{eq:a_i_def}) resemble Equations (\ref{eq:dw1}) and (\ref{eq:dw2}) for the trivial case of a pair of allies. Additionally, the constraint $0.0 < \mu \leq 0.50$ imposed in Equation (\ref{eq:update}) is also imposed in Equations (\ref{eq:dw1})--(\ref{eq:dw2}).

The popular DeGroot \cite{degroot1974reaching} and Hegselmann-Krause \cite{hegselmann2002opinion} models average the opinions of an agent and their allies, similar to Equation (\ref{eq:update_diff}) which averages the opinion differences between an agent and their allies and opponents. A common method of incorporating the influence of the media in deterministic models is to let the media be a special agent whose opinion never changes \cite{martins2010mass, sirbu2013opinion, pineda2015, brooks2020model}. In this paper, in contrast, the media are modelled as a random external signal with an intrinsic bias; they are not an agent in the network. Fang et al.'s model \cite{fang2019social}, one of the few opinion dynamics models  to treat probabilistic beliefs, also incorporates random external signals. It differs from this paper and Ref. \cite{nlow1} by excluding antagonistic relationships.

In the idealized scenario of inferring the bias of a coin, there are two possible external signals, heads or tails. In a real application involving media bias, there may be multiple external signals. As one idealized example, a newspaper's editorials may espouse liberal, middle-of-the-road, or conservative economic policies, corresponding to one bias parameter (degree of conservatism) and three possible external signals (the three policy types). Multiple bias parameters are also possible, which combine to control multiple external signals in diverse areas of social and economic life. Despite the limitations of the biased coin, it has the advantage of making the interpretation of the numerical simulations intuitive and tractable, which is valuable in a first pass at the problem. In \ref{sec:die_bias}, we briefly explore the problem of inferring the bias of a six-sided die in networks of up to three agents by modifying Equation (\ref{eq:actual_bayes}) and confirm that the resulting behavior is similar qualitatively to the two-sided coin.

\subsection{Categories of long-term behavior including intermittency}
\label{subsec:behavior_categories}

Networked agents display three types of long-term behavior. Stable agents do not change their mind once they make it up, i.e.\ $x_i(t,\theta)$ does not change significantly (by more than a user-selected tolerance) for $t > t_{\rm a}$, where $t_{\rm a}$ is termed the asymptotic learning time. By contrast, ``turbulent''\footnote{The term ``turbulent'' attracts the same meaning as in Ref. \cite{nlow1}. We do not use the alternative adjective ``chaotic'', because it implies sensitive dependence on initial conditions, which has not been verified rigorously yet \cite{nlow1}.} agents cannot make up their mind about $\theta$, i.e.\ $x_i(t,\theta)$ does not settle down to a fixed functional form for all $t$ but continues to evolve erratically or in an organized fashion (e.g.\ switching in a limit cycle between two beliefs). Finally, intermittent agents cycle between eras of stability and turbulence, where each era lasts for many time steps in general. A turbulent agent can be thought as a special case of an intermittent agent with short dwell times. The PDF of agent $i$ is stable for $\tau$ time steps if the following condition holds

\begin{equation}
\label{eq:asymp}
    \max_\theta | x_i(t',\theta)-x_i(t,\theta) | < \delta \max_\theta x_i(t, \theta) 
\end{equation}

\noindent for all $t \leq t' \leq t + \tau$ and $ 0\leq \theta \leq 1$, where $\delta=0.01$ is an analyst-selected tolerance.

\begin{figure}[h!tb]
    \centering
    \includegraphics[width=\linewidth]{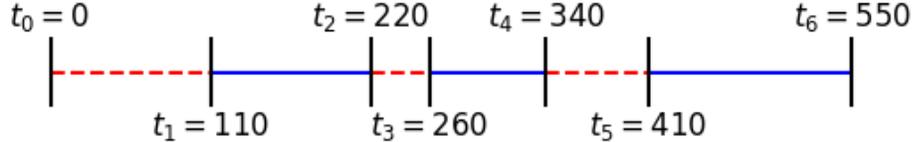}
    \caption{Defining stable and turbulent dwell times: a toy scenario of an agent with eras where Equation (\ref{eq:asymp}) is true (solid blue line) and false (dashed red line). In this toy scenario, the agent enters their first era of stability at $t_1$. We have a stable dwell time of $\stableDwell=t_2-t_1 > 100$. The agent then enters an era of turbulence at $t_2$. The turbulent dwell time stretches until the next era of stability exceeds 100 time steps. Hence, we have a turbulent dwell time of $\turbulentDwell=t_5-t_2$ instead of $t_3-t_2$. Similarly, the next stable dwell time is $\stableDwell=t_6-t_5 > 100$ instead of $t_4-t_3$.}
    \label{fig:example_category}
\end{figure}

Equation (\ref{eq:asymp}) suggests some key observables. Firstly, as noted above, if an agent is stable, we define the asymptotic learning time $\asympTime$ to be the smallest $t$ for which Equation (\ref{eq:asymp}) holds for $t<t'<T$, where $T$ is the number of time steps the automaton is run. Second, if we have an intermittent agent, we can define the (i) stable dwell time $\stableDwell$ and (ii) turbulent dwell time $\turbulentDwell$. The stable dwell time $\stableDwell$ is defined to be the number of time steps that an era of stability lasts, i.e.\ the number of consecutive time steps in which Equation (\ref{eq:asymp}) is true. A formal, unambiguous definition of $\stableDwell$ is presented in \ref{sec:ts_definition} with the help of a pseudocode snippet to assist the reader in reproducing the results in Sections \ref{sec:conditions} and \ref{sec:dwell_time_stats}.  Additionally, we impose an arbitrary threshold $\stableDwell \geq 100$ to distinguish intermittent behavior from turbulent behavior. We then define the turbulent dwell time $\turbulentDwell$ to be the number of time steps between the end of one stable era and the start of the next stable era. Figure \ref{fig:example_category} illustrates how we define $\stableDwell$ and $\turbulentDwell$ with reference to a toy scenario.

\section{Conditions for intermittency}
\label{sec:conditions}

Large networks are typically made up of smaller subnetworks. Therefore, in order to identify the conditions for intermittency in large networks, we first consider the smallest subnetworks that exhibit the phenomenon, namely those with $n = 4$. We identify three distinct types of intermittency in networks with $n=4$ in Section \ref{sec:conditions_n4}, deduce their implications for the structure of general networks (including those with $n>4$) exhibiting intermittency in Section \ref{sec:implication_condition}, and discuss the number of intermittent agents in larger networks (e.g.\ with $n=100$) in Section \ref{sec:num_intermittent}.

\subsection{Subnetworks with $n\leq 4$}
\label{sec:conditions_n4}

\begin{figure}[h!t]
    \centering
    \begin{subfigure}[b]{0.24\linewidth}
        \centering
        \includegraphics[width=\linewidth]{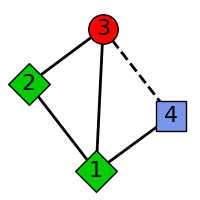}
        \caption{}
        \label{subfig:four_nodes_a}
    \end{subfigure}
    \begin{subfigure}[b]{0.24\linewidth}
        \centering
        \includegraphics[width=\linewidth]{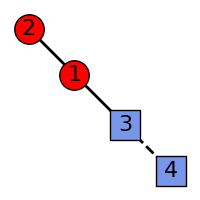}
        \caption{}
        \label{subfig:four_nodes_b}
    \end{subfigure}
    \begin{subfigure}[b]{0.24\linewidth}
        \centering
        \includegraphics[width=\linewidth]{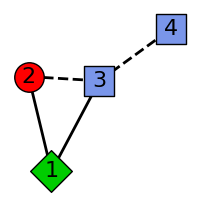}
        \caption{}
        \label{subfig:four_nodes_c}
    \end{subfigure}
    \caption{Intermittency in $n=4$ networks: three examples out of the 15 networks that display intermittency (out of 41 $n=4$ networks in total). The red circles, blue squares and green diamonds represent intermittent, stable and turbulent agents respectively. The solid lines between agents represents allies, while the dashed lines represent opponents. The networks in (a), (b) and (c) display Type A, B and C intermittency respectively. The other 12 networks with $n=4$ that display intermittency are depicted in Figure \ref{fig:four_nodes_gallery}.}
    \label{fig:four_nodes}
\end{figure}

We first generate all possible connected, non-isomorphic networks of size $n=3$ and $n=4$ with $A_{ij}=1,-1,0$. Here, connected means that no agent or group of agents are isolated, while two networks are isomorphic if we can obtain one by relabelling the agents of the other. We discard all allies-only or opponents-only networks, as they do not give rise to intermittent behavior as shown in previous work \cite{nlow1}. We are left with three $n=3$ networks and 41 $n=4$ networks. For each of these networks, we run the automaton $10^4$ times with randomized priors and $T=10^5$ coin tosses. We find that none of the $n=3$ networks display intermittency, in accord with previous work \cite{nlow1}. For $n=4$ networks, 15 out of the 41 possible networks display intermittency, three of which are illustrated in Figure \ref{fig:four_nodes}. For completeness, the other 12 intermittent networks are shown in \ref{sec:intermittent_gallery}.

\begin{figure}[h!t]
    \centering
    \begin{subfigure}[b]{0.49\linewidth}
        \centering
        \includegraphics[width=\linewidth]{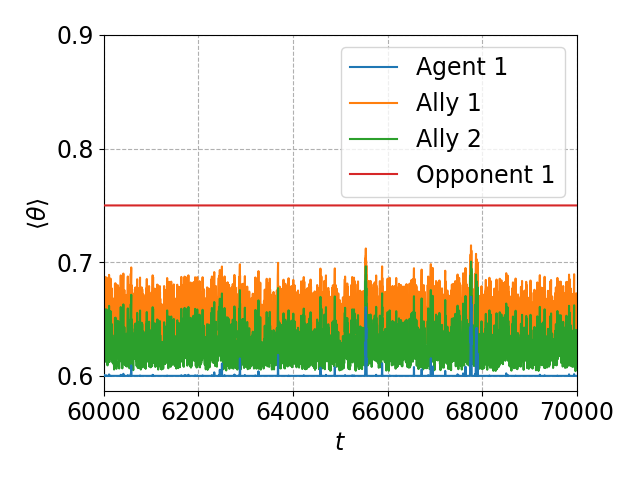}
        \caption{}
        \label{subfig:first_example_mean}
    \end{subfigure}
    \begin{subfigure}[b]{0.49\linewidth}
        \centering
        \includegraphics[width=\linewidth]{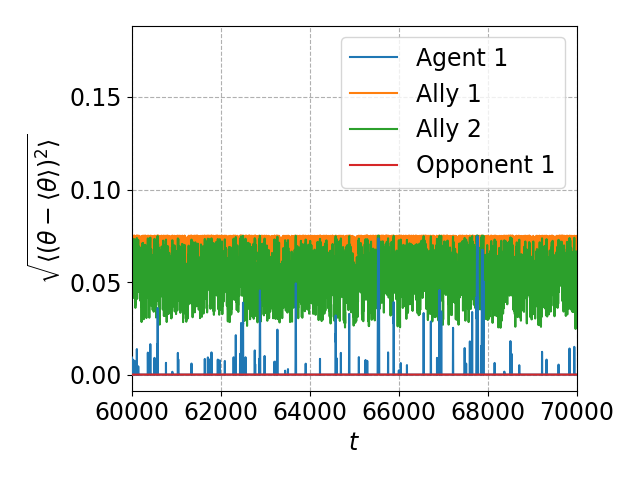}
        \caption{}
        \label{subfig:first_example_stddev}
    \end{subfigure}
    \begin{subfigure}[b]{0.49\linewidth}
        \centering
        \includegraphics[width=\linewidth]{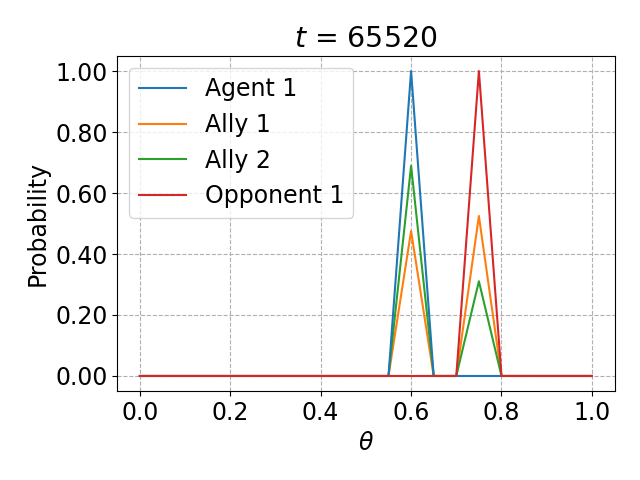}
        \caption{}
        \label{subfig:first_example_prior_stable}
    \end{subfigure}
    \begin{subfigure}[b]{0.49\linewidth}
        \centering
        \includegraphics[width=\linewidth]{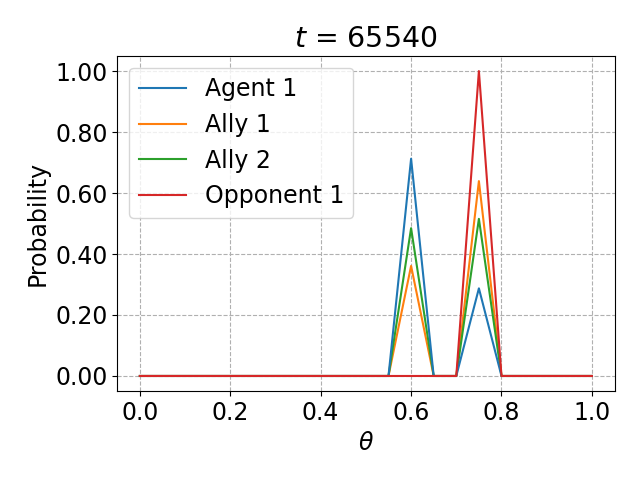}
        \caption{}
        \label{subfig:first_example_prior_unstable}
    \end{subfigure}
    \caption{Intermittency in the network shown in Figure \ref{subfig:four_nodes_a}: evolution of (a) $\mean$ and (b) $\stddev$ of an intermittent agent (agent 1, blue curve), its opponent (red curve) and two allies (green and orange curves) at $6\times 10^4 \leq t \leq 7 \times 10^4$. Snapshots of the beliefs $x_i(t,\theta)$ of the four agents at two particular times, during which the intermittent agent is in an era of temporary (c) stability and (d) turbulence.}
    \label{fig:four_nodes_first_example}
\end{figure}

What exactly causes intermittency to surface? For $n=4$ networks, we identify three distinct scenarios exemplified by the networks in Figure \ref{fig:four_nodes}. We label the three scenarios as Type A, B and C. The network shown in Figure \ref{subfig:four_nodes_a}, in which the one intermittent agent has two turbulent allies and one stable opponent, exhibits Type A intermittency. Figure \ref{fig:four_nodes_first_example} shows the evolution of $\mean$ and $\stddev$ of the four agents in Figures \ref{subfig:first_example_mean} and \ref{subfig:first_example_stddev} respectively, along with snapshots of their beliefs $x_i(t,\theta)$ at $t=65520$ (Figure \ref{subfig:first_example_prior_stable}) and $t=65540$ (Figure \ref{subfig:first_example_prior_unstable}), when the intermittent agent is temporarily stable and turbulent respectively. The source of Type A intermittency can be seen in Figures \ref{subfig:first_example_prior_stable} and  \ref{subfig:first_example_prior_unstable}. During the eras of stability, like in Figure \ref{subfig:first_example_prior_stable}, the intermittent agent 1 is fully confident in $\theta=0.60$, with $x_3(t,\theta=0.60)=1$ and zero for all other $\theta$. The intermittent agent has a stable opponent who is fully confident in $\theta=0.75$, and two turbulent allies with fluctuating confidence in $\theta=0.60,0.75$. Due to the opponent's high confidence in $\theta=0.75$ compared to the two allies, the intermittent agent is ``locked out'' from $\theta=0.75$. More precisely, because the strength of the internal signals is proportional to the difference in beliefs, as specified in Equation (\ref{eq:update}), we have $x_{\rm Ally 1}(t,\theta=0.75) + x_{\rm Ally 2}(t,\theta=0.75) < x_{\rm Opponent 1}(t,\theta=0.75) = 1$ at $t=65520$. However, purely due to chance, a run of heads starting at $t=65515$ boosts the confidence of the two allies in the belief $\theta=0.75$, whereupon they overpower the dissenting influence of the opponent, which causes the intermittent agent to suddenly gain confidence in $\theta=0.75$, as seen in Figure \ref{subfig:first_example_prior_unstable}. The intermittent agent enters a brief era of turbulence at $t=65523$, with fluctuating confidence in $\theta=0.60,0.75$ like its turbulent allies. This temporary turbulence persists until enough tails are observed to drive $x_3(t,\theta=0.75)$ to zero again and enter the next period of stability. The intermittency endures throughout the simulation, even when it is extended from $T=10^5$ to $T=10^7$ coin tosses.

Figure \ref{subfig:four_nodes_b} leads to a different scenario, which we label as Type B intermittency: the intermittent agents are not ``locked out'' of any $\theta$, unlike Type A intermittency in Figure \ref{subfig:four_nodes_a}. In Figure \ref{subfig:four_nodes_b}, the beliefs of the stable agents 3 and 4 are singly peaked at $\theta=0.4$ and $\theta=0.6$ respectively, with $x_3(t, \theta=0.4)=1$ and $x_4(t, \theta=0.6)=1$. The intermittent agents 1 and 2 have nonzero beliefs at both $\theta=0.4$ and $\theta=0.6$. The intermittent agents face two competing converging forces: agent 3 attempts to convince its ally agent 1, and by extension agent 2, that $\theta_0$ is 0.4 and not 0.6, while the coin tosses imply that $\theta_0$ is 0.6 and not 0.4. During the stable era of both agents 1 and 2, the converging influence of agent 3 overpowers the coin tosses to cause the beliefs of agent 1 to be singly peaked at $\theta=0.4$, with $x_1(t,\theta=0.4) \approx 1$ and $x_1(t,\theta=0.6) \ll 1$. When there is a run of heads, the coin tosses temporarily dominate, which boosts agent 1's confidence in $\theta=0.6$ sufficiently for agent 1 to enter an era of turbulence. Eventually, agent 3's influence overwhelms the coin tosses again, which returns agent 1 to another era of stability. Similarly to the network shown in Figure \ref{subfig:four_nodes_a}, intermittency endures throughout the simulation even for $T=10^7$ coin tosses.

A third mode of intermittency, which we label Type C, occurs in Figure \ref{subfig:four_nodes_c}, where the intermittency is driven mainly by the coin tosses, unlike Types A and B. In Figure \ref{subfig:four_nodes_c}, the intermittent agent has one turbulent ally and one stable opponent. Antagonistic interactions nullify the beliefs of the intermittent agent except at $\theta=0.50$ and $\theta=0.70$, while its opponent settles stably at $\theta=\theta_0=0.6$ with $x(t,\theta=\theta_0)=1$. In the short run, the binomial distribution is roughly equal at $\theta=0.50$ and $\theta=0.70$ for $\theta_0=0.60$. Suppose that during an era of stability, the beliefs of the intermittent agent are singly peaked at $\theta=0.50$. A run of heads boosts the intermittent agent's confidence in $\theta=0.70$. The intermittent agent enters an era of turbulence as it chooses between $\theta=0.50,$ $0.70$. The intermittent agent then enters the next era of stability when it observes sufficient coin tosses to settle on one of the two options $\theta=0.50$ and $\theta=0.70$. The cycling between $\theta=0.50$ and $\theta=0.70$ persists for a while, because those coin biases are not ``locked out'' by the opponents. Hence the intermittent agent never truly rules out either $\theta=0.50$ or $\theta = 0.70$. Contrary to the networks in Figures \ref{subfig:four_nodes_a} and \ref{subfig:four_nodes_b}, the intermittent behavior in Figure \ref{subfig:four_nodes_c} does not persist indefinitely. Upon extending the simulation from $T=10^5$ to $T=10^7$, the intermittent agent becomes stable at $t \approx 2 \times 10^5$ time steps. The intermittent behaviour disappears because the binomial distribution is higher at $\theta=0.50$ than at $\theta=0.70$ for $\theta_0=0.60$, i.e.\ the coin tosses favor $\theta=0.50$ over $\theta=0.70$ in the long run. The intermittency does not resurface because machine precision rounds $x_i(t,\theta=0.70)$ to zero which prevents Equation (\ref{eq:bayes}) from increasing $x_i(t,\theta=0.70)$, and there are no allies with nonzero confidence in $\theta=0.70$, which prevents Equation (\ref{eq:update}) from increasing $x_i(t,\theta=0.70)$.

The three types of intermittency identified above may not exhaust all possibilities, e.g.\ $n=5$ networks may have intermittency which cannot be traced back to Types A, B, and C. However, $n=5$ networks are too numerous to examine manually, and an automated tracker of the patterns like in Figure \ref{fig:four_nodes} currently consumes too much computational time.

\subsection{Implications for network structure}
\label{sec:implication_condition}

We can deduce some general properties of the network structure required to yield each of the three modes of intermittency in Figure \ref{fig:four_nodes}. 

(i) In Figure \ref{subfig:four_nodes_a}, Type A intermittency occurs when the allies of the intermittent agent are sufficiently confident to overcome the antagonism of the opponent. We require the intermittent agent to have at least two allies; the beliefs of the stable opponent are singly peaked with $x_{\rm Opponent}(t, \theta = \meanSub{\rm Opponent}) = 1$, so a lone ally can never have sufficient confidence to increase the intermittent agent's confidence in $\theta = \meanSub{\rm Opponent}$ by Equation (\ref{eq:update}), because one has $x_i(t,\theta) \leq 1$ for all $\theta$. Furthermore, the two allies must experience turbulent nonconvergence, i.e.\ have fluctuating beliefs, in order to ``switch'' the turbulent agent between eras of stability and turbulence. None of the other intermittent $n=4$ networks shown in \ref{sec:intermittent_gallery} features an intermittent agent with two turbulent allies and one stable opponent. 

(ii) In Figure \ref{subfig:four_nodes_b}, the intermittent agent 1 has a stable ally that settles at $\theta \neq \theta_0$. Type B Intermittency occurs because agent 1 receives conflicting information from its ally who promotes $\theta \neq \theta_0$, and the coin tosses which promote $\theta=\theta_0$. We can deduce that the intermittent ally must have a stable ally, who must have at least one opponent who settles at $\theta=\theta_0$. The opponent is required to lock the ally out of $\theta=\theta_0$, which implies that this opponent cannot be antagonistic with the intermittent agent (otherwise the intermittent agent would also be locked out of $\theta=\theta_0$). Two networks, shown in Figures \ref{subfig:four_nodes_extra_a} and \ref{subfig:four_nodes_extra_b}, display the same Type B intermittent behavior as in Figure \ref{subfig:four_nodes_b} and also have an intermittent agent who is allied to a stable agent who is an opponent to another stable agent. 

(iii) Type C intermittency occurs in Figure \ref{subfig:four_nodes_c}, in which the intermittent agent has nonzero beliefs at two $\theta$ values straddling $\theta_0$. In the short run, the binomial distributions at these two $\theta$ values are roughly equal, i.e.\ an isolated agent would not favor one of these $\theta$ values over the other. Being locked out of $\theta=\theta_0$ implies that the intermittent agent must have an opponent. Additionally, an ally is required to convince the intermittent agent to have nonzero beliefs straddling $\theta_0$, otherwise the opponent would ``lock out'' all $\theta > \theta_0$ or $\theta < \theta_0$. The remaining ten networks shown in Figures \ref{subfig:four_nodes_extra_c}--\ref{subfig:four_nodes_extra_l} display similar Type C intermittent behaviors as shown in Figure \ref{subfig:four_nodes_c}. The networks, except Figure \ref{subfig:four_nodes_extra_g}, all have a common feature: at least one intermittent agent is embedded in a $n=3$ subnetwork containing two alliances and one antagonism. Such a $n=3$ subnetwork is called an unbalanced triad according to the social science theory of structural balance \cite{heider1946attitudes, davis1967clustering} and is recognized as an instigator of opinion instability in political and historical contexts \cite{antal2006balance, pratto2014}.

The above network properties are insufficient conditions for intermittency. Changing the sequence of coin tosses, while keeping the priors and $A_{ij}$ unchanged, does not always produce intermittency. Out of the $10^4$ simulations performed on the network shown in Figure \ref{subfig:four_nodes_a}, in which the coin tosses and priors are randomized, about 25$\%$ yield intermittent behavior. By contrast, for each of the other 14 networks, less that $1\%$ of the realizations display intermittency. The possibility exists, therefore, that other $n=4$ networks also produce intermittency, albeit rarely. This implies that, although intermittency only occurs in certain networks, the coin tosses (i.e.\ published media products) play a significant role in determining if intermittent behavior surfaces in practice in any individual historical situation.

\subsection{Number of intermittent agents}
\label{sec:num_intermittent}

Out of the 15 $n=4$ networks with intermittency, nine of them have one intermittent agent (one Type A, two Type B, six Type C), five have two intermittent agents (one Type B, four Type C) while the final network (Type C) has three intermittent agents. How many intermittent agents do larger networks have? To answer this question, we study \barabasi networks with $n=100$ and attachment parameter $m=3$, as defined in the \texttt{Python} package \texttt{networkx} \cite{networkx}, as a representative example. The \barabasi model \cite{barabasi1999emergence} generates networks that are approximately scale-free. We study \barabasi networks because many networks in the real world are thought to be approximately scale-free \cite{barabasi2003scale} as well as to facilitate comparison with the results in Ref. \cite{nlow1}. The original \barabasi model does not distinguish between alliances and antagonisms. To obtain a mix of allies and opponents, we randomly choose a percentage $q$ of all links in the networks to be opponents, while the other $1-q$ of the links are allies. More details about how to construct larger networks are given in Ref. \cite{nlow1}. We vary $q$ in the range $0.01 \leq q \leq 0.99$. For each unique value of $q$, we run the automaton with $T=10^6$ coin tosses until we have 100 simulations with at least one intermittent agent per simulation.

\begin{figure}[h!t]
    \centering
    \begin{subfigure}[b]{0.75\linewidth}
        \centering
        \includegraphics[width=\linewidth]{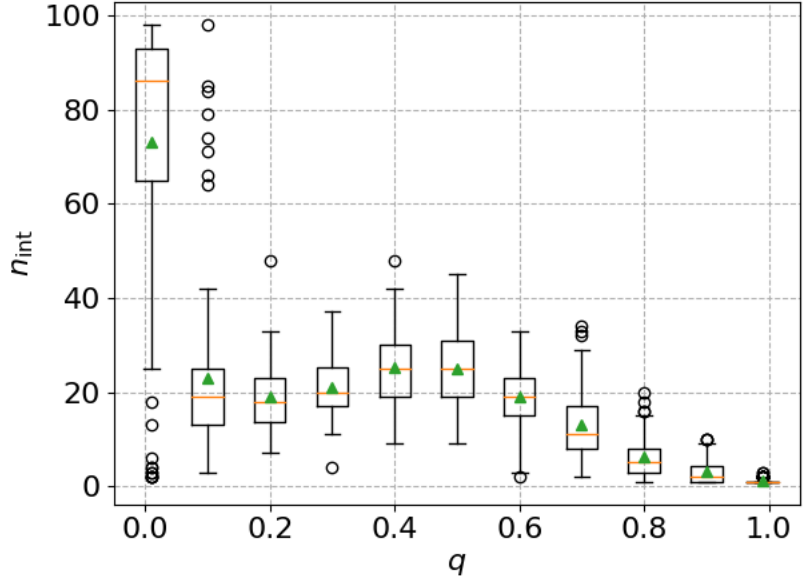}
        \caption{}
    \end{subfigure}
    \caption{Boxplot of the number of intermittent agents, $\numInt$, versus the percentage of antagonistic links, $q$. For each $q$, we run 100 independent simulations with randomized priors and $T=10^6$ coin tosses, with $A_{ij}$ generated using the \barabasi model with $n=100$ and attachment parameter $m=3$ \cite{networkx}. The orange line and green triangle within each box represents the median and mean of the data set respectively. The box extends to the first and third quartiles, while the whiskers extend to one and a half times the interquartile range from the edges of the box. Outlier simulations beyond the whiskers are drawn separately as open circles for emphasis. The first and third quartiles, mean and median of $\numInt$ for $q=0.99$ all equal unity.}
    \label{fig:num_int}
\end{figure}

Figure \ref{fig:num_int} shows the number of intermittent agents, $\numInt$, as a function of $q$. The highest mean ($\langle \numInt \rangle = 73$) and median (median $\numInt =86$) occur at $q=0.01$. The relatively high $\numInt$ at $q=0.01$ compared to other $q$ values is likely caused by a ``domino'' effect: one intermittent agent exerts their converging influence across its allies, who then drive intermittency across their own allies. At $0.10 \leq q \leq 0.50$, the mean and median $\numInt$ are roughly constant. However, $q=0.10$ contains eight outlier simulations with $\numInt > 60$, which is higher than $\max \numInt = 48$ for $0.20 \leq q \leq 0.50$. For $q > 0.50$, $\numInt$ tends to unity, as $q$ tends to unity, with $85\%$ of the $q=0.99$ simulations having only one intermittent agent. The $q=0.99$ simulations have a maximum of three intermittent agents, which reflects the fact that every $q=0.99$ network has three alliances --- an intermittent agent can drag along two allies at most. Consistent with previous work \cite{nlow1}, we find that intermittent agents always have at least one ally, and an agent with only antagonistic links is never intermittent.

\section{Dwell time statistics}
\label{sec:dwell_time_stats}

In Section \ref{sec:conditions}, we study what kind of networks yield intermittent agents. In this section, we quantify the temporal patterns in the three distinct kinds of intermittency identified in Section \ref{sec:conditions}. Specifically, we study the statistics of their dwell times, $\stableDwell$ and $\turbulentDwell$ as defined in Section \ref{subsec:behavior_categories}. We denote the histograms of $\stableDwell$ and $\turbulentDwell$ to be $\stablePDF$ and $\turbulentPDF$ respectively. Importantly for practical applications, it turns out that the different kinds of intermittency produce different dwell time statistics. Turning this around, one can use the observed dwell time statistics to infer the structure of a real-world network, at least in principle.

\subsection{Subnetworks with $n=4$}
\label{sec:dwell_small}

\begin{figure}[h!t]
    \centering
    \begin{subfigure}[b]{0.87\linewidth}
        \centering
        \includegraphics[width=0.49\linewidth]{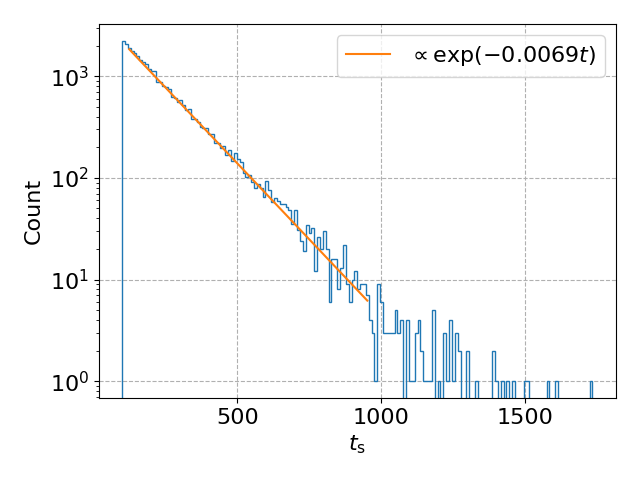}
        \includegraphics[width=0.49\linewidth]{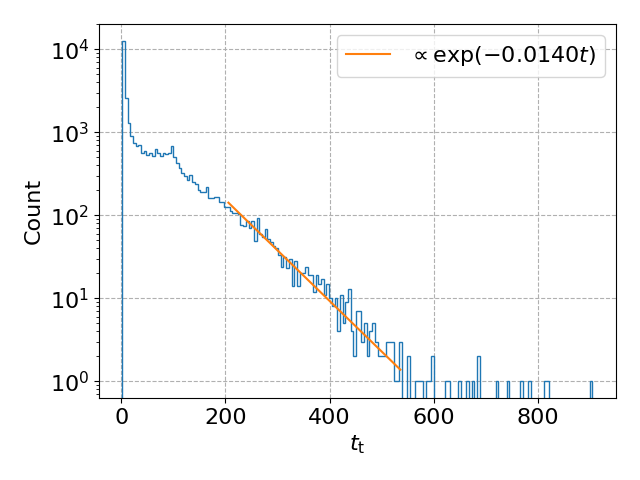}
        \caption{}
        \label{subfig:dwell_1}
    \end{subfigure}
    \begin{subfigure}[b]{0.87\linewidth}
        \centering
        \includegraphics[width=0.49\linewidth]{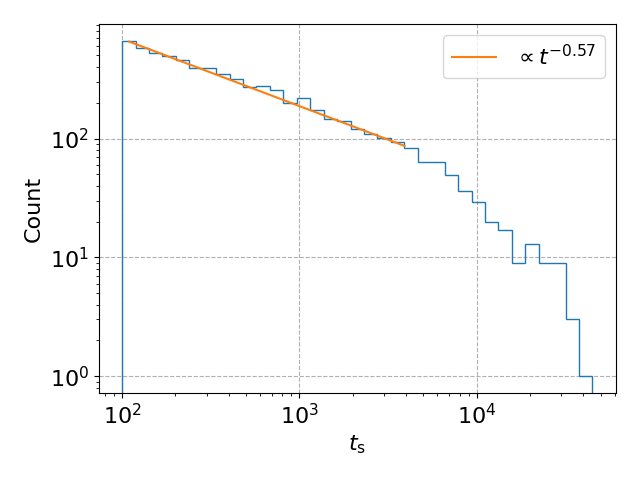}
        \includegraphics[width=0.49\linewidth]{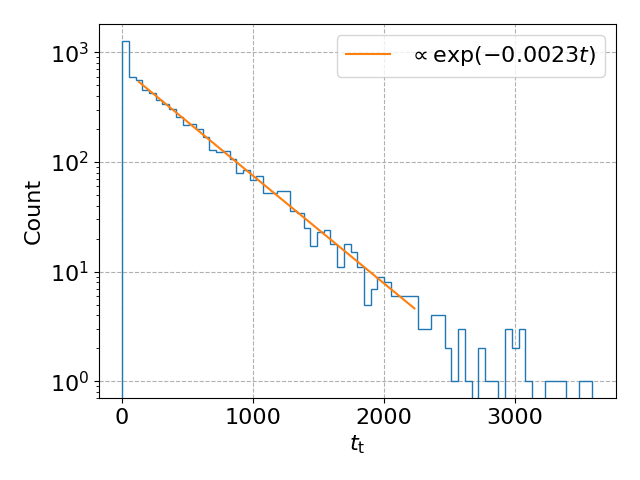}
        \caption{}
        \label{subfig:dwell_9}
    \end{subfigure}
    \begin{subfigure}[b]{0.87\linewidth}
        \centering
        \includegraphics[width=0.49\linewidth]{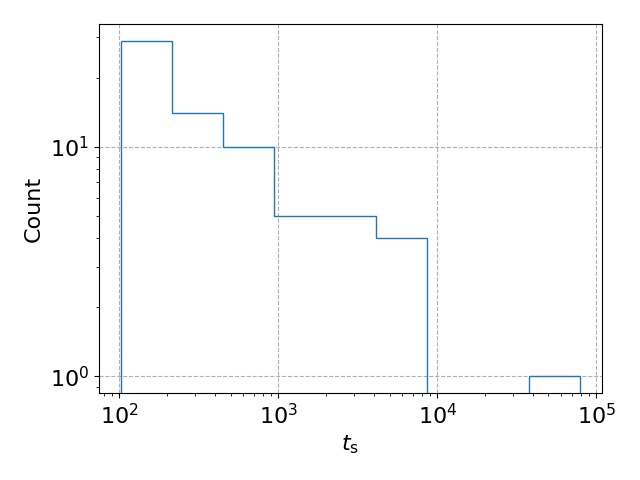}
        \includegraphics[width=0.49\linewidth]{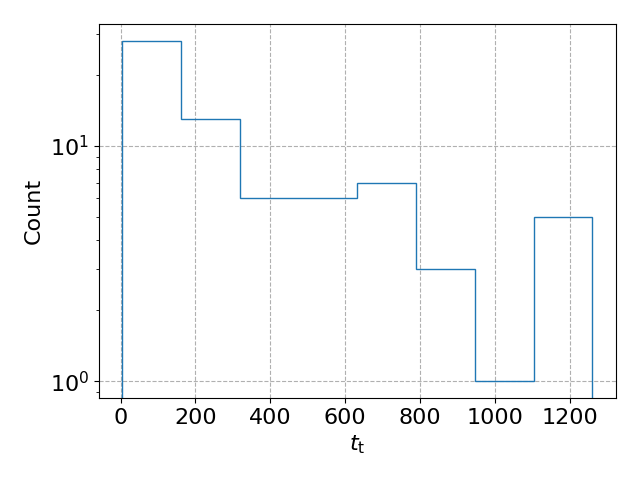}
        \caption{}
        \label{subfig:dwell_14}
    \end{subfigure}
    \caption{Histograms of the dwell times $\stableDwell$ and $\turbulentDwell$ for the intermittent agents in (a) Figure \ref{subfig:four_nodes_a}, (b) Figure \ref{subfig:four_nodes_b} (agent 1) and (c) Figure \ref{subfig:four_nodes_c} in a particular realization with $T=10^7$ coin tosses. For reference, the best fit exponential and power law are overplotted in (a) and (b) as orange curves. The intermittent behavior disappears for $t\gtrsim 10^5$ in (c), so there are fewer instances of $\stableDwell$ and $\turbulentDwell$ compared to (a) and (b).}
    \label{fig:dwell_stats}
\end{figure}

Figures \ref{subfig:dwell_1}, \ref{subfig:dwell_9} and \ref{subfig:dwell_14} shows the histograms of $\stablePDF$ and $\turbulentPDF$ of the intermittent agents in Figures \ref{subfig:four_nodes_a}, \ref{subfig:four_nodes_b} and \ref{subfig:four_nodes_c} respectively. In Figure \ref{subfig:dwell_1}, which displays Type A intermittency, we see that $\stablePDF$ follows an exponential distribution, while $\turbulentPDF$ approximately follows an exponential distribution for $\turbulentDwell \gtrsim 2 \times 10^2$. By contrast, in Figure \ref{subfig:dwell_9}, which displays Type B intermittency, $\stablePDF$ approximately follows a power law for $\stableDwell \lesssim 5 \times 10^3$, while $\turbulentPDF$ follows an exponential. We find that the two networks shown in Figures \ref{subfig:four_nodes_extra_a} and \ref{subfig:four_nodes_extra_b}, which exhibit Type B intermittency, have similar shapes of $\stablePDF$ and $\turbulentPDF$ as shown in Figure \ref{subfig:dwell_9}. Hence, by observing the dwell time statistics of the intermittent agent, we may be able to infer if the agent is embedded in a network similar to the networks shown in Figures \ref{subfig:four_nodes_a} and \ref{subfig:four_nodes_b}. We are unable to gather sufficient instances of $\stableDwell$ and $\turbulentDwell$ for the network shown in Figure \ref{subfig:four_nodes_c} to meaningfully discuss its dwell time statistics, because the intermittent behavior disappears after $\sim 10^5$ time steps.

The characteristic time-scales for $\stableDwell$ and $\turbulentDwell$ are different. For example, Figure \ref{subfig:dwell_1} shows that $\turbulentDwell$ has a shorter timescale than $\stableDwell$, with exponents $\approx -0.014 \turbulentDwell$ and $-0.0069 \stableDwell$ respectively. This means that the intermittent agent generally spends less time in a turbulent state than in a stable state.

\begin{figure}[h!t]
    \centering
    \begin{subfigure}[b]{0.49\linewidth}
        \centering
        \includegraphics[width=\linewidth]{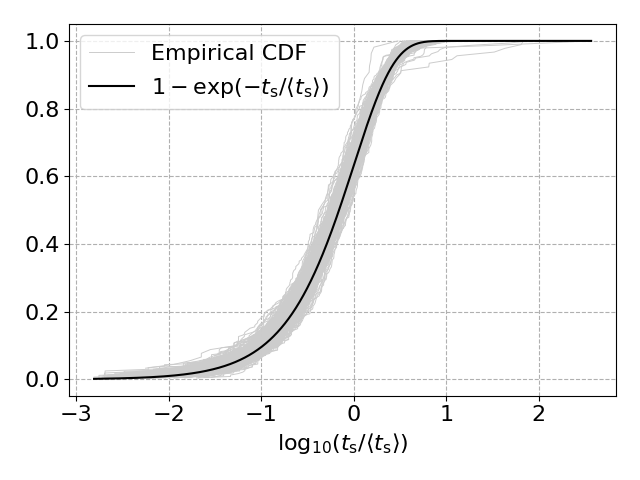}
        \caption{}
        \label{subfig:exp_four_cdf}
    \end{subfigure}
    \begin{subfigure}[b]{0.49\linewidth}
        \centering
        \includegraphics[width=\linewidth]{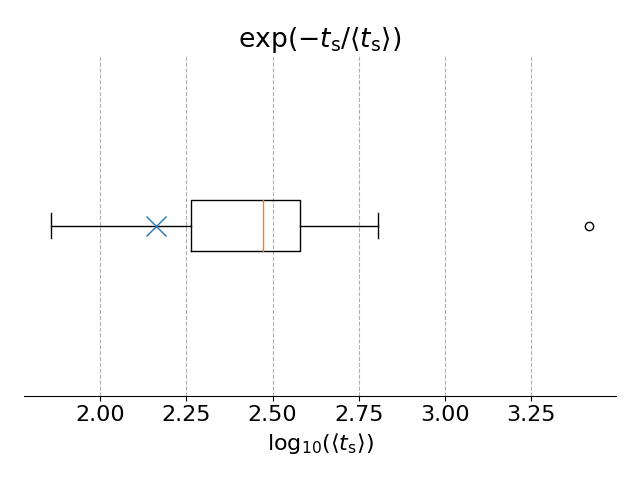}
        \caption{}
        \label{subfig:exp_four}
    \end{subfigure}
    \caption{Dwell time distributions for 259 realizations of the network shown in Figure \ref{subfig:four_nodes_a}. (a) Empirical cumulative distribution function (CDF) of $\stableDwell / \stableMean$ for the 259 dwell time distributions, overplotted on each other as light gray curves. For reference, the function $1-\exp(\stableDwell / \stableMean)$ is also plotted as a solid black curve. (b) Boxplot of the exponents of the 259 dwell time distributions $\stablePDF$. The boxplot is defined in Figure \ref{fig:num_int}. For reference, a blue cross marks the exponent of $\stablePDF$ shown in Figure \ref{subfig:dwell_1}. The characteristic time-scale of $\stablePDF$ is denoted as $\stableMean$ as estimated (with maximum likelihood) from the sample mean. \cite{walpole2012probability}.}
    \label{fig:dwell_exp_four_nodes}
\end{figure}

As mentioned in Section \ref{sec:implication_condition}, intermittency occurs in about $25\%$ of the $10^4$ simulations performed on the network shown in Figure \ref{subfig:four_nodes_a} with randomized priors and coin tosses. Figure \ref{subfig:exp_four_cdf} shows the empirical cumulative distribution functions of $\stableDwell / \stableMean$ for each intermittent agent, overplotted on each other as light gray staircases, while Figure \ref{subfig:exp_four} shows a boxplot of $\stableMean$ for the network shown in Figure \ref{subfig:four_nodes_a}. Figure \ref{subfig:exp_four_cdf} shows that every cumulative distribution function of $\stableDwell / \stableMean$ follows the curve $1-\exp(-\stableDwell/\stableMean)$, so every $\stablePDF$ can be approximated by an exponential function. The first, second and third quartiles of the exponent of $\stablePDF$ are $-0.0055 \stableDwell$, $-0.0034 \stableDwell$ and $-0.0026 \stableDwell$ respectively. A similar boxplot for the exponents of $\turbulentPDF$ cannot be generated because Figure \ref{fig:dwell_exp_four_nodes} shows that $\turbulentPDF$ is exponential only for a certain range of $\turbulentDwell$, and an automated identification of this range is currently unavailable. Similarly, boxplots of $\stableDwell$ and $\turbulentDwell$ for Types B and C intermittency cannot be provided due to the rarity of the occurrence of intermittency in these networks, as mentioned in Section \ref{sec:implication_condition}.

\subsection{Larger networks}
\label{sec:dwell_big}

Do the characteristics of the dwell times differ for larger networks? Using the same set of simulations that yield Figure \ref{fig:num_int} in Section \ref{sec:num_intermittent}, we investigate the dwell time statistics of \barabasi networks with $n=100$, whose percentage of antagonistic links $q$ falls in the range $0.01 \leq q \leq 0.99$. In Section \ref{sec:dwell_small}, we find that Type A intermittency has exponential $\stablePDF$, while Type B intermittency has exponential $\turbulentPDF$. In this section, we determine how many intermittent agents in the \barabasi networks have exponential $\stablePDF$ and $\turbulentPDF$, using the same set of simulations that yield Figure \ref{fig:num_int}.

To determine if $\stableDwell$ or $\turbulentDwell$ are exponentially distributed, we first bin the dwell times using the \texttt{histogram} function in the \texttt{Python} package \texttt{numpy} \cite{numpy} with the parameter \texttt{bins='auto'}. We then fit a straight line with the least squares method using logarithmically-scaled bin counts to obtain an estimate of the exponent. The distribution of the dwell time is considered exponential if the coefficient of determination statistic $R^2$ \cite{walpole2012probability} exceeds a tolerance, which we take to be 0.9 arbitrarily. We aim to classify the dwell time distributions approximately. Future work will test for more complicated histogram shapes, such as a broken power law for certain ranges of $\stableDwell$ as seen in Figure \ref{subfig:dwell_9}.

\begin{figure}[h!t]
    \centering
    \begin{subfigure}[b]{0.49\linewidth}
        \centering
        \includegraphics[width=\linewidth]{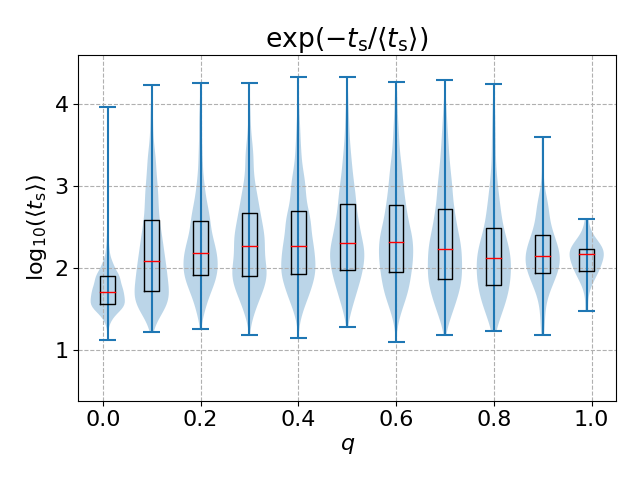}
        \caption{}
        \label{subfig:exponent_s}
    \end{subfigure}
    \begin{subfigure}[b]{0.49\linewidth}
        \centering
        \includegraphics[width=\linewidth]{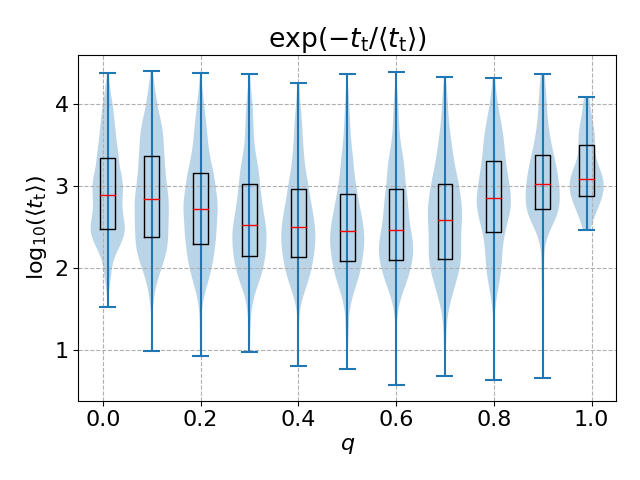}
        \caption{}
        \label{subfig:exponent_t}
    \end{subfigure}
    \begin{subfigure}[b]{0.49\linewidth}
        \centering
        \includegraphics[width=\linewidth]{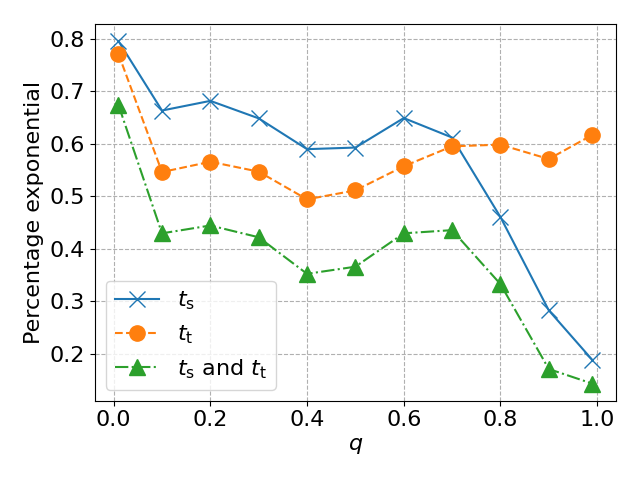}
        \caption{}
        \label{subfig:exponent_percent}
    \end{subfigure}
    \begin{subfigure}[b]{0.49\linewidth}
        \centering
        \includegraphics[width=\linewidth]{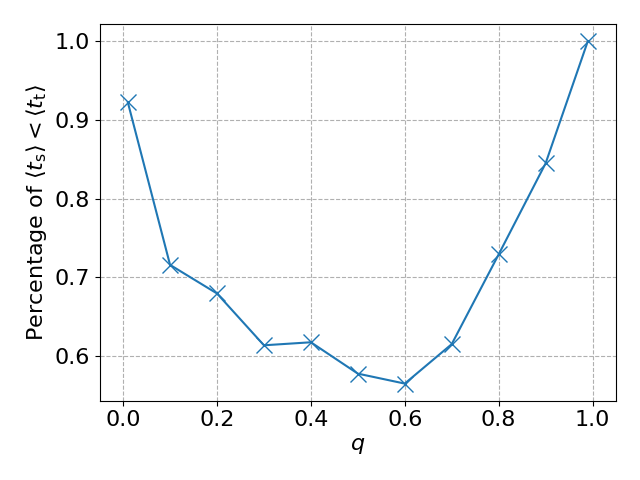}
        \caption{}
        \label{subfig:exponent_s_bigger}
    \end{subfigure}
    \caption{Dwell time distributions. (a) and (b) Violin plots of the exponents of (a) $\stablePDF$ and (b) $\turbulentPDF$ when they are found to be exponential, as functions of the percentage of antagonistic links $q$. Each panel shows the empirical probability distribution on its side colored in blue, with short blue horizontal line segments drawn at the extrema. The boxes represent the first and third quartiles, while the red line inside the boxes represents the median. (c) Percentage of intermittent agents with exponential PDFs $\stablePDF$ and $\turbulentPDF$ represented as solid blue and dashed orange curves respectively, while the dotted-dashed green curve represents the percentage of intermittent agents that have both exponential $\stablePDF$ and $\turbulentPDF$ simultaneously, as functions of $q$. (d) Percentage of simulations with $\stableMean < \turbulentMean$ when both are exponential, as a function of $q$. The plots in this figure are generated using the same set of simulations that yield Figure \ref{fig:num_int}.}
    \label{fig:exponents}
\end{figure}

Figures \ref{subfig:exponent_s} and \ref{subfig:exponent_t} show the exponents of $\stablePDF$ and $\turbulentPDF$ that are found to be exponential, while Figure \ref{subfig:exponent_percent} shows the percentage of intermittent agents that have exponential dwell times as a function of $q$. When both $\stableDwell$ and $\turbulentDwell$ have exponential distributions, Figure \ref{subfig:exponent_s_bigger} shows the percentage of realizations with $\stableMean < \turbulentMean$. In general, we find $\stableMean < \turbulentMean$ when $\stablePDF$ and $\turbulentPDF$ are both exponential, with $100\%$ of the $q=0.99$ simulations having $\stableMean < \turbulentMean$. Moreover, $\stableDwell$ has the shortest timescale for $q=0.01$, with a median $\stablePDF$ exponent of $-10^{-1.7} \stableDwell$, compared to between $-10^{-2.1} \stableDwell$ and $-10^{-2.3} \stableDwell$ for all $q>0.01$. For $\turbulentDwell$, the timescales lengthen, as $q$ approaches zero or unity, with the shortest median timescale occurring at $q=0.50$. Figure \ref{subfig:exponent_percent} shows that there is a higher percentage of exponential $\stablePDF$ the $\turbulentPDF$ for $q \leq 0.70$, while the opposite is true for $q > 0.70$. We find that $q=0.01$ yields the highest percentage of exponential $\stablePDF$ and $\turbulentPDF$. For $0.10 \leq q \leq 0.70$, the percentage of exponential $\stablePDF$ is roughly constant with $q$, while the percentage drops for $q>0.70$. By contrast, the percentage of exponential $\turbulentPDF$ is roughly constant for all $q\geq 0.10$, although there appears to be a slight upwards trend for $q \geq 0.40$.

Preliminary visual inspections reveals that some of the dwell time distributions are poorly approximated by an exponential or a power law. Three example outliers are shown in \ref{sec:weird_dwell}. When the outliers are log-binned, the $\stableDwell$ histogram decreases with increasing $\stableDwell$ for small $\stableDwell$ and has a bump (Figure \ref{subfig:dwell_weird_1}) or a few scattered events (Figure \ref{subfig:dwell_weird_2}) at large $\stableDwell$. A $\turbulentDwell$ histogram (Figure \ref{subfig:dwell_weird_3}) has a bump at small $\turbulentDwell$, with a smaller secondary bump at large $\turbulentDwell$.

The above findings beget many subtle questions. Why are most dwell time distributions exponential? What kind of networks produce non-exponential dwell time distributions? Why are the timescales for $\stableDwell$ generally shorter than $\turbulentDwell$? We are not in a position to answer these questions theoretically from first principles in this introductory, empirical paper. We encourage future theoretical investigations along these lines, while acknowledging the serious challenges for analytic calculations presented by far-from-equilibrium feedback phenomena in complex networks.

\section{Conclusion}

Networked consumers of media products form political opinions (e.g.\ about bias in newspaper editorials) through a combination of independent observation and peer pressure from political allies and opponents in the network. Multi-agent simulations reveal several counterintuitive phenomena, one of which is intermittency. Intermittent agents cycle between eras of stability, when their beliefs are constant for many time steps, and turbulence, when their beliefs fluctuate from one time step to the next. In this paper, we map the real-world problem of assessing media bias onto the idealized problem of assessing the bias of a coin. We analyze networks of Bayesian learners with unsure beliefs, which are updated by Bayes's rule upon observing the coin toss, modified by a form of ``peer pressure'' mediated by a mix of supportive and antagonistic interactions with allies and opponents. Using multi-agent simulations, we investigate every connected, non-isomorphic $n \leq 4$ network to determine what kind of network structure yields intermittent behavior. We also determine the number of intermittent agents in \barabasi networks with $n=100$ as a function of the percentage of antagonistic links in the network, $q$. We use \barabasi networks for concreteness, but qualitatively similar results are found for \erdos and square lattice networks; see Section 7 in Ref. \cite{nlow1} for more details. Finally, we calculate the dwell time statistics of the intermittent networks with $n \leq 4$  and $n=100$.

The main findings of the paper are as follows. (i) None of the $n=3$ networks and 15 out of the 41 $n=4$ networks display intermittency. (ii) The $n=4$ networks display three different types of intermittent behavior. Type A intermittency is caused by a run of heads (or tails) which gives the allies of the intermittent agent sufficient confidence to overcome the antagonism of the opponent of the intermittent agent. Type B intermittency surfaces, when the intermittent agent has a stable ally who settles on the wrong coin bias, which subjects the intermittent agent to two contradictory converging influences, namely the coin tosses and the wrong ally. Type C intermittency occurs when the agent is locked into two possible coin biases straddling the true coin bias, which are approximately equally likely in the short run. (iii) Types A and B persist for $t \geq 10^7$ time steps and arguably indefinitely, while Type C disappears after $\sim 10^5$ time steps, once the intermittent agent settles on one option. (iv) For Type A intermittency, $\stablePDF$ is exponential for all $\stableDwell$, and its exponent has first, second and third quartiles of $-0.0055 \stableDwell$, $-0.0034 \stableDwell$ and $-0.0026 \stableDwell$ respectively, while $\turbulentPDF$ is exponential for a certain $\turbulentDwell$ range. Type B produces a broken power law $\stablePDF$, while $\turbulentPDF$ is exponential for all $\turbulentDwell$. (v) The shapes of the dwell time distributions differ for Types A and B, so one can potentially use observations of the dwell times to infer the network structure, at least in principle. (vi) For \barabasi networks with $n=100$ and randomized priors, coin tosses and network structures $A_{ij}$, more than $50\%$ of the dwell time distributions are found to be exponential for each $0.01 \leq q \leq 0.99$. The exception is $\stablePDF$ for $q \geq 0.80$, in which the percentage of exponential $\stablePDF$ dips below $50\%$ and continues to decrease as $q$ approaches unity. The highest percentages (nearly $80\%$) of exponential $\stablePDF$ and $\turbulentPDF$ occur at $q=0.01$.

The opinion dynamics in this paper is highly idealized in at least four ways. (i) It does not incorporate the political psychology of individuals. (ii) It models the peer pressure between individuals in a limited fashion through $A_{ij}$ and Equations (\ref{eq:update})--(\ref{eq:a_i_def}). (iii) It evolves opinions based on a Bayesian update rule which is broadly plausible but has not been tested systematically in real-life experiments. (iv) It considers $A_{ij}$ to be static, even though real-life networks evolve. Nevertheless, with due reserve, we advance for discussion some possible social implications, which arguably flow from the model. First and foremost is the observation that ``changing one's mind'' occurs even in a static network; an individual may vacillate in their perceptions of media bias, driven by peer pressure rather than individual psychology, even if their connections with political allies and opponents do not change with time. The influence of the social environment on opinion stability --- as distinct from opinion formation --- has received some attention in the social science literature \cite{hoffman2007}, but more work needs to be done, especially with respect to media bias. Second, the dwell times associated with intermittency are $\gtrsim 10^2$ times longer typically than the time between the arrival of new pieces of information (e.g.\ coin tosses). This is important: intermittency is a metastable phenomenon; it is not the same as randomly changing one's mind, every time a new datum arrives. Taken at face value, the idealized results in Section \ref{sec:dwell_time_stats} imply that a person may hold a fixed opinion for (say) a year about the bias of a newspaper that issues daily editorials (coin tosses) then transition to a state of uncertainty which also lasts (say) a year, when their opinion fluctuates frequently. Third, because only certain network structures yield certain intermittent behavior (Types A--C), we may use observations of Type A--C behavior (including dwell time distributions) to infer the underlying network structure in real-life applications, at least in principle. This is potentially important in situations where the network structure is hard to measure independently (e.g.\ political alliances), but the temporal patterns of opinions (e.g.\ dwell times) are easier to measure with standard (e.g.\ longitudinal) surveys. Finally, increasing the percentage of antagonistic links in the network leads to fewer intermittent agents, which may imply that too much internal dissent within a community solidifies people's opinions and reduces vacillation, at least as far as the peer-pressure-driven dynamics are concerned. We emphasize again that peer pressure is only one factor among many, and it is modeled in a highly idealized fashion in this paper. Individual psychology plays a key role too, and it is excluded entirely from the automaton in Section \ref{sec:model}.

In addition to its limitations from a social science perspective, this paper can also be extended in several technical directions, which we will investigate in future work. For example, we only investigate the structure of $n \leq 4$ networks to test for intermittency. Hence, as noted in Section \ref{sec:conditions}, there is no guarantee that $n \geq 5$ networks are limited to displaying intermittency of Types A, B, and C. Other types of intermittency may exist, driven by other subnetworks with $n\geq 5$. Moreover, the results for \barabasi networks with $n=100$ should be generalized to other scale-free networks. Ref. \cite{nlow1} finds that many of the qualitative features of the media bias problem are unchanged in \erdos and square lattice networks, but the specific results in this paper on the conditions for intermittency (Section \ref{sec:conditions}) and dwell time statistics (Section \ref{sec:dwell_time_stats}) should be checked in their own right.

\section*{Acknowledgements}

Parts of this research are supported by the University of Melbourne Science Graduate Scholarship - 2020, the Australian Research Council (ARC) Centre of Excellence for Gravitational Wave Discovery (OzGrav) (project number CE170100004) and ARC Discovery Project DP170103625. 

\bibliographystyle{elsarticle-num}
\bibliography{ref.bib}

\appendix

\section{Bayes's rule and its role in evolving beliefs and inferring the bias of a coin}
\label{sec:bayes_detailed}

Bayes's rule is a way of combining the information gained from observations of a system together with preexisting knowledge of the system \cite{Johnson2022-nm}. In the context of Equations (\ref{eq:bayes})--(\ref{eq:a_i_def}), an agent's preexisting knowledge about the coin bias is given by $x_i(t=0, \theta)$. Upon observing a coin toss, the agent gains new information about $\theta_0$. This new information is expressed mathematically as $P[S(t)|\theta]$ given in Equation (\ref{eq:actual_bayes}). For example, upon observing a heads, Equations (\ref{eq:bayes}) and (\ref{eq:actual_bayes}) tell the agent that a coin bias of $\theta=0.0$ is impossible because $\theta=0.0$ corresponds to a coin that only returns tails.

To illustrate how Bayes's rule operates when inferring the bias of a coin (as an idealized analogy for media bias), we solve the problem analytically in the special case where there is no peer pressure among agents in the network, i.e. we neglect temporarily the complications introduced by Equations (\ref{eq:update})--(\ref{eq:a_i_def}) for illustrative purposes. Specifically, we show below that the iterated application of Bayes's rule, as specified by Equations (\ref{eq:bayes}) and (\ref{eq:actual_bayes}), reproduces the familiar binomial formula for the probability of a sequence of coin tosses. The aim of the exercise is to help orient the reader with respect to the role played by Bayes's rule in this paper. It is not a substitute for a thorough discussion of Bayes's rule, which can be found in many classic textbooks, e.g. Ref. \cite{Gelman2013-vu}.

Without internal signals, we may relabel $x'_i(t+1/2,\theta)$ as $x_i(t+1,\theta)$ (one-step update rule) and rewrite Bayes's rule in Equation (\ref{eq:bayes}) as

\begin{equation}
    \label{eq:bayes_rewritten}
    x_i(t+1,\theta) = \frac{P[S(t)|\theta]}{P[S(t)]}  x_i(t,\theta).
\end{equation}

\noindent Equation (\ref{eq:bayes_rewritten}) does not remember time ordering when it is applied repeatedly; if an agent observes a heads followed by a tails, Equation (\ref{eq:bayes_rewritten}) updates their beliefs in exactly the same way as if they observe a tails followed by a heads. This is because (i) the probability of returning a heads at time $t$ is independent of the probability of returning a heads at time $t+1$, as implied by Equation (\ref{eq:actual_bayes}), and (ii) Bayes's rule updates the beliefs multiplicatively, so if two observations are made, incorporating the first observation then the second is mathematically identical to incorporating the second observation then the first.

Recall that $P[S(t)|\theta]$ is the probability that the coin returns the signal $S(t) \in \{\text{heads},$ $\text{tails}\}$, assuming that the coin has a bias $\theta$. If $T$ coin tosses return $T_{\rm heads}$ heads, then $\prod^{T-1}_{t=0} P[S(t)|\theta]$ is given by the $T$-fold product of Equation (\ref{eq:actual_bayes}) with itself, i.e. $T_{\rm heads}$ factors of $\theta$ and $T-T_{\rm heads}$ factors of $1-\theta$. Normalizing the result, we obtain

\begin{equation}
    x_i(t=T,\theta) = x_i(t=0,\theta) \prod_{t=0}^{T-1} \frac{P[S(t)|\theta]}{P[S(t)]}
\end{equation}
    
\noindent with 

\begin{equation}
\label{eq:bayes_binom}
    \prod_{t=0}^{T-1} P[S(t)|\theta] = \frac{T!}{T_{\rm heads}!(T-T_{\rm heads})!} \theta^{T_{\rm heads}} (1-\theta)^{T-T_{\rm heads}}.
\end{equation}

Equation (\ref{eq:bayes_binom}) is the familiar binomial distribution, as expected for $T$ tosses of a coin. In the limit $T \rightarrow \infty$, one finds $\prod_{t=0}^{T-1} P[S(t)|\theta=\theta_0] \rightarrow 1$ and $\prod_{t=0}^{T-1} P[S(t)|\theta \neq \theta_0] \rightarrow 0$. Therefore, assuming that the agent starts with a nonzero belief at the true coin bias $\theta_0$, i.e. $x_i(t=0,\theta = \theta_0) \neq 0$, then they are able to correctly infer the coin bias.

\section{Visualization of prior generation and the discretization of the coin bias $\theta$}
\label{sec:discretization_visual}

\begin{figure}[h!tb]
    \centering
    \includegraphics[width=0.5\linewidth]{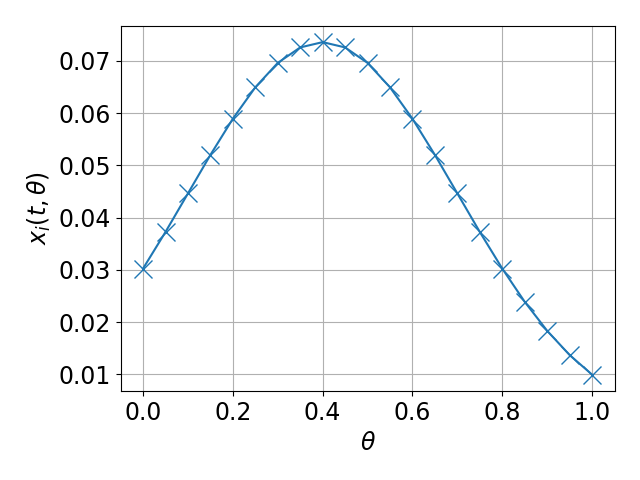}
    \caption{A prior generated by sampling a Gaussian with mean 0.4 and standard deviation 0.3 before it is truncated to the domain $0.0 \leq \theta \leq 1.0$. The crosses indicate that the Gaussian has been sampled at $\theta=\{0.00, 0.05, ..., 1.00\}$. This figure is reproduced from Ref. \cite{nlow1}.}
    \label{fig:prior_generation}
\end{figure}

In the model described in Section \ref{sec:update_rules}, $0.0 \leq \theta \leq 1.0$ is discretized into 21 regularly-spaced values. Following Ref. \cite{nlow1}, the priors of each agent are randomly generated by sampling Gaussians with means and standard deviations in the ranges [0.0, 1.0] and [0.1, 0.4] respectively in the domain $0.0 \leq \theta \leq 1.0$. Figure \ref{fig:prior_generation} shows an example of how a prior is generated, and how $\theta$ is discretized.

\section{Inferring the bias of a six-sided die}
\label{sec:die_bias}

In the model presented in this paper, we consider the problem of inferring the bias of a coin. A coin only returns two different external signals, which is overly restrictive in many practical applications concerning media bias, as discussed in Section \ref{sec:alternative_update_rules}. In this appendix, we briefly explore the problem of inferring the bias of a six-sided die to show how the model described in Section \ref{sec:update_rules} can be extended to incorporate more external signals. We find that the behavior of the six-sided die and two-sided coin are qualitatively similar, although of course there are differences in the details.

Suppose a six-sided die has a bias $\theta$ and returns six different signals, $S(t) = \{1,2,3,4,5,6\}$. We define the likelihood of each signal as 

\begin{equation}
\label{eq:die_likelihood}
    P[S(t)|\theta] =
    \begin{dcases}
        \left(1 + \theta \right)/6,& \quad \text{if } S(t) = 1 \\
        \left(1 - \theta \right)/6,& \quad \text{if } S(t) = 2 \\
        \left(1 + \theta^2 \right)/6,& \quad \text{if } S(t) = 3 \\
        \left(1 - \theta^2 \right)/6,& \quad \text{if } S(t) = 4 \\
        \left(1 + \theta^3 \right)/6,& \quad \text{if } S(t) = 5 \\
        \left(1 - \theta^3 \right)/6,& \quad \text{if } S(t) = 6
    \end{dcases}
\end{equation}

\noindent with $-1.0 \leq \theta \leq 1.0$ to ensure $0 \leq P[S(t),\theta] \leq 1$. The bias $\theta$ may be interpreted as how much the die favors odd numbers, where a bias of $\theta=0.0$ indicates a fair die. For any $0.0 < | \theta | < 1.0$, the likelihoods $P[S(t)=1,2 | \theta]$ deviate the most from a fair die, while $P[S(t)=5,6 | \theta]$ deviate the least from a fair die. 

To facilitate comparison with a biased coin \cite{nlow1}, we only consider three simple networks: a pair of allies ($A_{12}=1$), a pair of opponents ($A_{12}=-1$) and a triad with internal tensions ($A_{12}=-1$, $A_{23}=A_{31}=1$). We run the automaton described in Section \ref{sec:update_rules} once for each of the three networks, with $T=10^4$ coin tosses. We discretize $\theta$ into 21 evenly-spaced values, i.e. $\theta = \{-1.0, -0.9,..., 0.9, 1.0\}$. For the pair of allies and pair of opponents, $x_1(t=0,\theta)$ and $x_2(t=0,\theta)$ are generated by sampling Gaussians with means 0.4 and $-0.2$, and standard deviations 0.3 ad 0.4 respectively. For the triad, the priors of each agent are generated by sampling Gaussians with means and standard deviations in the ranges [$-1.0$, 1.0] and [0.1, 0.8] respectively. We set $\theta_0=0.1$ without loss of generality.

\begin{figure}[h!t]
    \centering
    \begin{subfigure}[b]{0.45\linewidth}
        \centering
        \includegraphics[width=\linewidth]{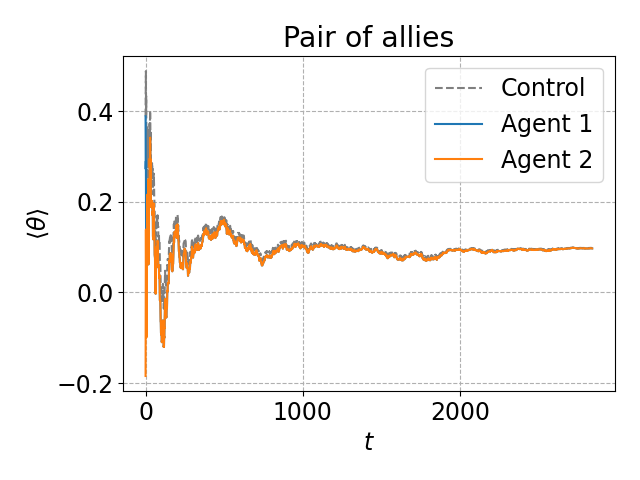}
        \caption{}
        \label{subfig:die_allies_mean}
    \end{subfigure}
    \begin{subfigure}[b]{0.45\linewidth}
        \centering
        \includegraphics[width=\linewidth]{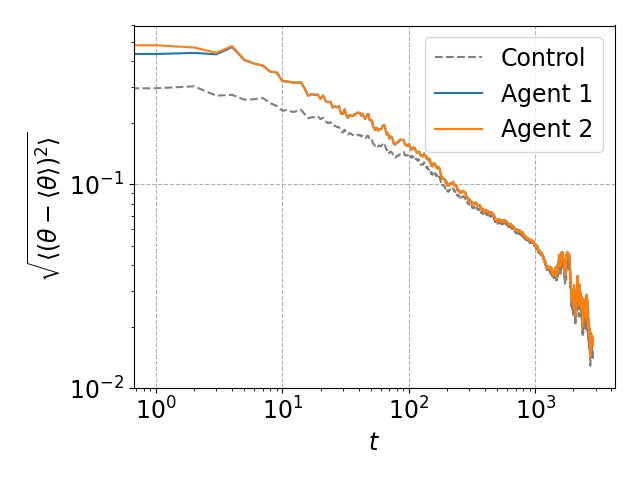}
        \caption{}
        \label{subfig:die_allies_stddev}
    \end{subfigure}
    \begin{subfigure}[b]{0.45\linewidth}
        \centering
        \includegraphics[width=\linewidth]{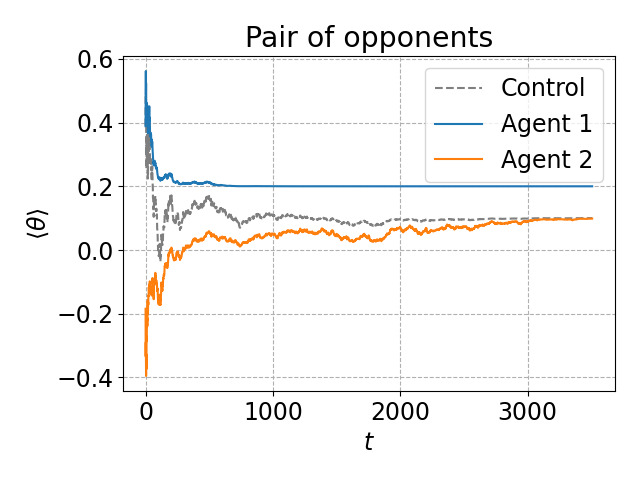}
        \caption{}
        \label{subfig:die_opponents_mean}
    \end{subfigure}
    \begin{subfigure}[b]{0.45\linewidth}
        \centering
        \includegraphics[width=\linewidth]{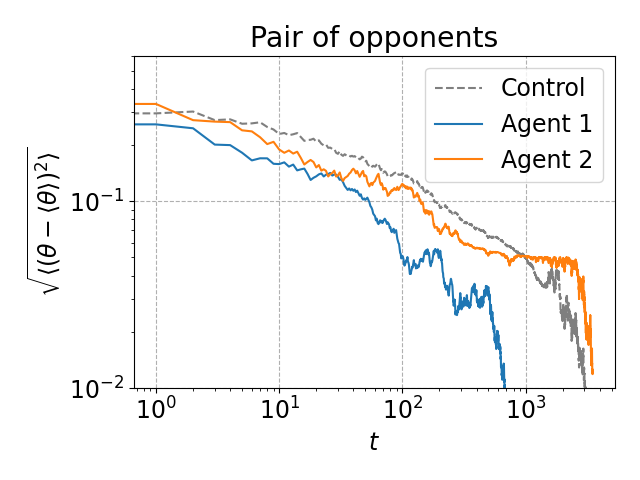}
        \caption{}
        \label{subfig:die_opponents_stddev}
    \end{subfigure}
    \begin{subfigure}[b]{0.45\linewidth}
        \centering
        \includegraphics[width=\linewidth]{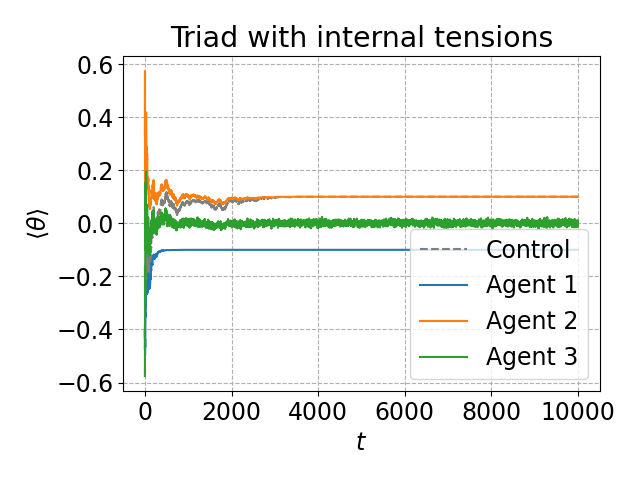}
        \caption{}
        \label{subfig:die_triad_mean}
    \end{subfigure}
    \begin{subfigure}[b]{0.45\linewidth}
        \centering
        \includegraphics[width=\linewidth]{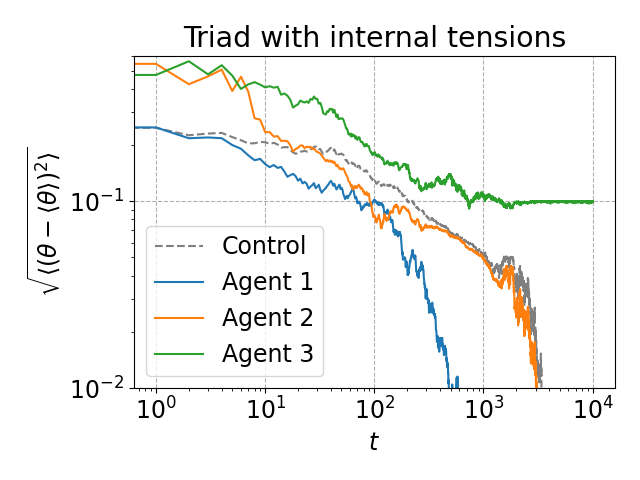}
        \caption{}
        \label{subfig:die_triad_stddev}
    \end{subfigure}
    \caption{Inferring the bias of a six-sided die: the evolution of the mean and standard deviation of the beliefs about $\theta$ from one particular simulation for (a) and (b) a pair of allies ($A_{12}=1$), (c) and (d) a pair of opponents ($A_{12}=1$) and (e) and (f) a triad with internal tensions ($A_{12}=-1$, $A_{23}=A_{31}=1$). The pair of allies and pair of opponents achieve asymptotic learning at $\asympTime = 2841$ and $\asympTime = 3508$ respectively, while the triad with internal tensions fails to achieve asymptotic learning for $t \leq T=10^4$. Parameters: $\theta_0=0.1$, $\mu=0.25$, $T=10^4$. Overall the behavior resembles qualitatively that of a two-sided die, as can be verified by comparing with Figures B.20, 3 and 8 in Ref. \cite{nlow1}.}
    \label{fig:die_sims}
\end{figure}

Figure \ref{fig:die_sims} shows the evolution of the mean beliefs, $\mean$ and their standard deviations, $\stddev$, from a particular simulation for each of the three networks: a pair of allies in Figures \ref{subfig:die_allies_mean} and \ref{subfig:die_allies_stddev}, a pair of opponents in Figures \ref{subfig:die_opponents_mean} and \ref{subfig:die_opponents_stddev}, and a triad with internal tensions in Figures \ref{subfig:die_triad_mean} and \ref{subfig:die_triad_stddev}. The sequence of coin tosses in all three simulations are the same. The pair of allies and pair of opponents achieve asymptotic learning, as defined in Section \ref{subsec:behavior_categories}, at $\asympTime = 2841$ and $\asympTime = 3508$ respectively. The triad with internal tensions does not achieve asymptotic learning within $T=10^4$ die observations.

Comparing to the results in Ref. \cite{nlow1}, the qualitative behaviours of the three networks shown in Figure \ref{fig:die_sims} for a six-sided die are similar to a two-sided coin. For the pair of allies, the beliefs of each agent converge onto each other (asymptotic learning), with $\meanSub{1} \approx \meanSub{2}$ and $\stddevSub{1} \approx \stddevSub{2}$. For the pair of opponents, agent 1 infers the wrong bias with $\meanSub{1} \rightarrow 0.2 \neq \theta_0$, while agent 2 infers the correct bias with $\meanSub{2} \rightarrow 0.1 = \theta_0$. Agent 1 becomes confident in the wrong bias at $t=t_{{\rm a}, 1}=838$, quicker compared to agent 2 who has $t_{{\rm a}, 2}=3508$. For the triad with internal tensions, agent 3 is allied to agents 1 and 2, but agents 1 and 2 are opponents. Agent 3 fails to achieve asymptotic learning because its two allies provide conflicting information, with $\meanSub{1} \rightarrow -0.1 \neq \theta_0$ and $\meanSub{2} \rightarrow 0.1 = \theta_0$. All three outcomes are the same as in Ref. \cite{nlow1}, including the counterintuitive outcome that agents in opposition often develop confidence in the wrong conclusion faster than the right conclusion.

As befits a preliminary check, this appendix compares the qualitative behaviour of networks with three agents or fewer in inferring the bias of a coin and a six-sided die. It is beyond the scope of this paper to check the qualitative behaviour of larger networks in inferring the bias of a six-sided die.

\section{Definition of the stable dwell time $t_{\rm s}$}
\label{sec:ts_definition}


The definition of the stable dwell time, $\stableDwell$, is hard to express compactly and unambiguously in words. In this appendix, we write down and justify a fragment of pseudocode for $\stableDwell$, in case the reader wishes to reproduce the results in Section \ref{sec:dwell_time_stats}. Listing \ref{code:ts} shows a \texttt{Python} pseudocode of how the stable dwell times $\stableDwell$ are calculated for each agent in the network if the agent exhibits intermittency. The pseudocode is commented to assist the reader.

\begin{listing}[!ht]
\begin{minted}
[
frame=lines,
framesep=2mm,
baselinestretch=1.2,
bgcolor=LightGray,
fontsize=\footnotesize,
linenos
]
{python}
for agent in list_of_agents:
  stable_start_time = 0  # tracks when the agent enters an era of stability
  current_ts = 0  # tracks how many time steps the agent is currently stable
  agent.list_of_ts = []  # stores the stable dwell times for this agent
  for t in range(1, T):  # T is the number of coin tosses
    if agent.is_eq6_true(t, stable_start_time):  # is eq (6) true?
      current_ts += 1
    else:  # eq (6) is false
      if current_ts >= 100:  # stable dwell times must exceed 100 time steps
        agent.list_of_ts.append(current_ts)
      current_ts = 0
      stable_start_time = t
\end{minted}
\caption{A \texttt{Python} pseudocode showing how the stable dwell time $\stableDwell$ is calculated for each agent in the network when the automaton is run for $T$ time steps. The list of $\stableDwell$ values for the agent is stored in \texttt{agent.list\_of\_ts}.}
\label{code:ts}
\end{listing}

\section{Gallery of intermittent $n=4$ networks}
\label{sec:intermittent_gallery}

\begin{figure}[h!t]
    \centering
    \begin{subfigure}[b]{0.19\linewidth}
        \centering
        \includegraphics[width=\linewidth]{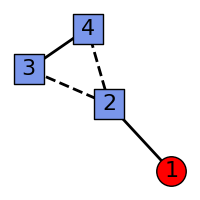}
        \caption{}
        \label{subfig:four_nodes_extra_a}
    \end{subfigure}
    \begin{subfigure}[b]{0.19\linewidth}
        \centering
        \includegraphics[width=\linewidth]{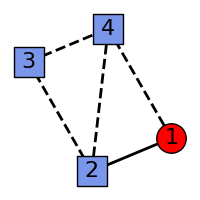}
        \caption{}
        \label{subfig:four_nodes_extra_b}
    \end{subfigure}
    \begin{subfigure}[b]{0.19\linewidth}
        \centering
        \includegraphics[width=\linewidth]{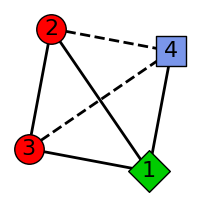}
        \caption{}
        \label{subfig:four_nodes_extra_c}
    \end{subfigure}
    \begin{subfigure}[b]{0.19\linewidth}
        \centering
        \includegraphics[width=\linewidth]{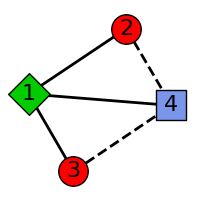}
        \caption{}
    \end{subfigure}
    \begin{subfigure}[b]{0.19\linewidth}
        \centering
        \includegraphics[width=\linewidth]{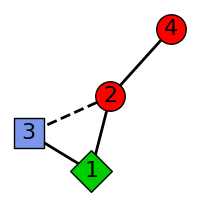}
        \caption{}
    \end{subfigure}
    \begin{subfigure}[b]{0.19\linewidth}
        \centering
        \includegraphics[width=\linewidth]{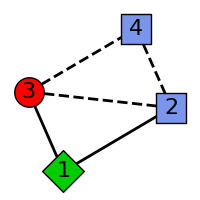}
        \caption{}
    \end{subfigure}
    \begin{subfigure}[b]{0.19\linewidth}
        \centering
        \includegraphics[width=\linewidth]{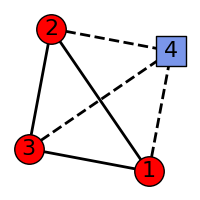}
        \caption{}
        \label{subfig:four_nodes_extra_g}
    \end{subfigure}
    \begin{subfigure}[b]{0.19\linewidth}
        \centering
        \includegraphics[width=\linewidth]{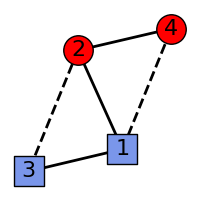}
        \caption{}
    \end{subfigure}
    \begin{subfigure}[b]{0.19\linewidth}
        \centering
        \includegraphics[width=\linewidth]{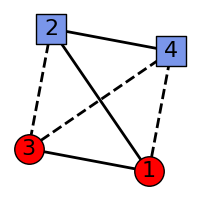}
        \caption{}
    \end{subfigure}
    \begin{subfigure}[b]{0.19\linewidth}
        \centering
        \includegraphics[width=\linewidth]{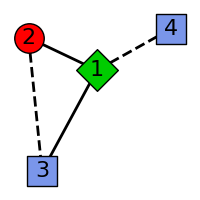}
        \caption{}
    \end{subfigure}
    \begin{subfigure}[b]{0.19\linewidth}
        \centering
        \includegraphics[width=\linewidth]{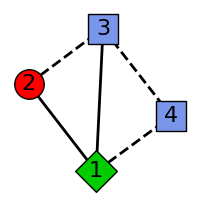}
        \caption{}
    \end{subfigure}
    \begin{subfigure}[b]{0.19\linewidth}
        \centering
        \includegraphics[width=\linewidth]{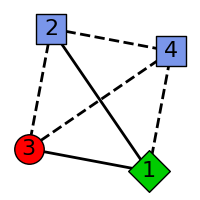}
        \caption{}
        \label{subfig:four_nodes_extra_l}
    \end{subfigure}
    \caption{Intermittency in $n=4$ networks: the 12 networks that display intermittency, other than the three shown in Figure \ref{fig:four_nodes}. The red circles, blue squares and green diamonds represent intermittent, stable and turbulent agents respectively. The solid lines between agents represents allies, while the dashed lines represent opponents. The two networks in (a) and (b) display Type B intermittency, while the other ten networks in (c)--(l) display Type C intermittency.}
    \label{fig:four_nodes_gallery}
\end{figure}

In Section \ref{sec:dwell_small}, we find that 15 out of 41 $n=4$ networks display intermittency. We visualize which agents are intermittent, stable and turbulent for three out of these 15 networks in Figure \ref{fig:four_nodes}. In Figure \ref{fig:four_nodes_gallery}, we visualize the remaining 12 networks. The networks in \ref{subfig:four_nodes_extra_a} and \ref{subfig:four_nodes_extra_b} display Type B intermittency, while the other ten networks display Type C intermittency.

\section{Dwell time histograms that are not exponentials or power laws}
\label{sec:weird_dwell}

\begin{figure}[h!t]
    \centering
    \begin{subfigure}[b]{0.49\linewidth}
        \centering
        \includegraphics[width=\linewidth]{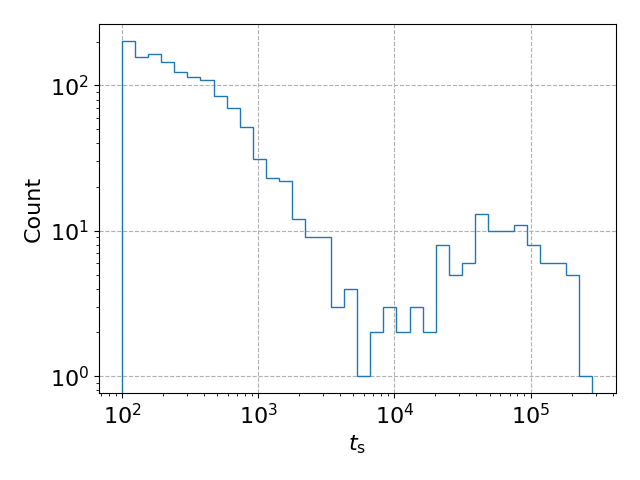}
        \caption{}
        \label{subfig:dwell_weird_1}
    \end{subfigure}
    \begin{subfigure}[b]{0.49\linewidth}
        \centering
        \includegraphics[width=\linewidth]{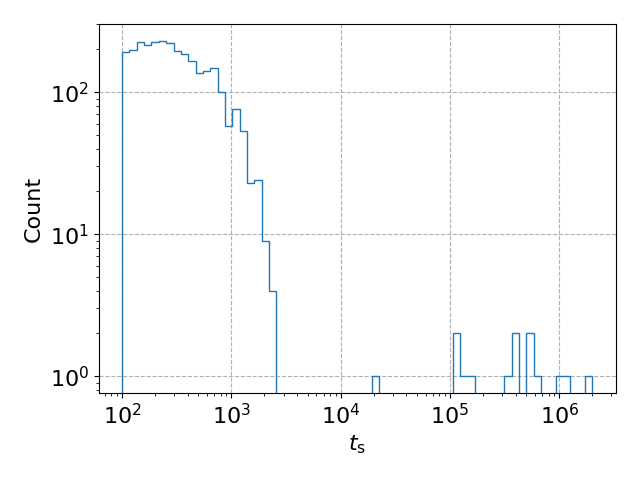}
        \caption{}
        \label{subfig:dwell_weird_2}
    \end{subfigure}
    \begin{subfigure}[b]{0.49\linewidth}
        \centering
        \includegraphics[width=\linewidth]{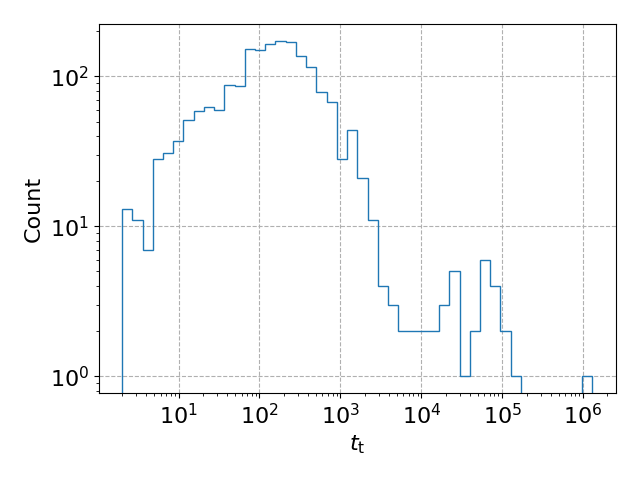}
        \caption{}
        \label{subfig:dwell_weird_3}
    \end{subfigure}
    \caption{Three examples of the dwell time statistics that do not approximately follow an exponential or power-law distribution. The examples are taken from a particular $q=0.50$ simulation with $T=10^7$ coin tosses. (a) and (b) $\stablePDF$ of two different intermittent agents. (c) $\turbulentPDF$ of another intermittent agent.}
    \label{fig:dwell_weird}
\end{figure}

In Section \ref{sec:dwell_small}, we find that the distributions of the dwell times $\stableDwell$ and $\turbulentDwell$ are well approximated by exponentials for Type A intermittency, and power laws and exponentials respectively for Type B. Preliminary visual checks of some of the dwell time distributions that are not classified as exponential (as described in Section \ref{sec:dwell_big}) in \barabasi networks with $n=100$ reveal that some of them are poorly approximated by exponentials and power laws. Figure \ref{fig:dwell_weird} shows three such examples where the dwell times are log-binned. Figure \ref{subfig:dwell_weird_1} shows that the $\stableDwell$ histogram decreases with increasing $\stableDwell$ for $\stableDwell \lesssim 10^4$ and has a bump at $\stableDwell \gtrsim 10^4$. Figure \ref{subfig:dwell_weird_2} is similar in shape to Figure \ref{subfig:dwell_weird_1} for $\stableDwell \lesssim 10^4$, but at $\stableDwell \gtrsim 10^4$ we instead see a few scattered events. Figure \ref{subfig:dwell_weird_3} shows a $\turbulentDwell$ histogram which has a primary bump at $\turbulentDwell \sim 10^2$, and a secondary bump at $\turbulentDwell \sim 10^5$.

\end{document}